\documentclass[reprint,superscriptaddress,nofootinbib,longbibliography]{revtex4-2}
\usepackage{graphicx} 
\usepackage{microtype}
\usepackage[english]{babel}
\usepackage[utf8]{inputenx}
\usepackage{wrapfig}
\usepackage{float}
\usepackage{ifthen,pifont,epsfig,latexsym,amsmath,amssymb}
\usepackage[strict]{changepage}
\usepackage{graphicx}
\usepackage{amsmath}
\usepackage{amsfonts}
\usepackage[T1]{fontenc}
\usepackage{listings}
\usepackage{braket}
\usepackage{xcolor}
\usepackage[margin=2cm]{geometry}
\usepackage{comment}
\usepackage{physics}
\usepackage{breqn}

\usepackage{hyperref}
 \hypersetup{
 colorlinks = true, 
 urlcolor = blue, 
 linkcolor = blue, 
 citecolor = red 
 }

\newcommand{\INFN}{INFN - Sezione collegata di Salerno, Complesso Univ. Monte S. Angelo, I-80126 Napoli, Italy}

\newcommand{\UNISA}{Physics Department ``E.R. Caianiello'', Universit\`a degli Studi di Salerno, Via Giovanni Paolo II, 132, I-84084 Fisciano (Sa), Italy}

\newcommand{\CNR}{CNR-SPIN, I-84084 Fisciano (Salerno), Italy, c/o Universit\`a di Salerno, I-84084 Fisciano (Salerno), Italy}

\newcommand{\UNINA}{CNR-SPIN and Physics Department ``Ettore Pancini'', Universit\`a degli Studi di Napoli Federico II, Via Cinthia, 80126-Napoli, Italy}

\begin{document}

\title{
Inhomogeneous superconductivity in (001), (110), and (111) KTaO$_3$
   two-dimensional electronic gases: effect of electronic confinement on the transition temperature
}

\author{Mattia Trama}
\email{mtrama@unisa.it}
\affiliation{\UNISA}
\affiliation{\INFN}

\author{Roberta Citro}
\affiliation{\UNISA}
\affiliation{\INFN}
\affiliation{\CNR}

\author{Carmine Antonio Perroni}
\email{carmine.perroni@unina.it}
\affiliation{\INFN}
\affiliation{\CNR}
\affiliation{\UNINA}

\begin{abstract}
Experimental observations on KTaO$_3$ superconductivity report widely different critical temperatures with a clear hierarchy among different crystallographic orientations. Accounting for the first time for spatial confinement, we connect the inhomogeneous superconducting order parameter with the spatial extent of the two-dimensional electron gas. We reproduce the observed hierarchy within a unified framework, based on a self-consistent tight-binding slab model with local spin-singlet $s$-wave pairing. The methods developed here can readily be extended to include more general interaction models, while still consistently including the characteristic spatial structure of the electron gas.
\end{abstract}

\maketitle

{\bf\textit{Introduction.}}---Two-dimensional electron gases (2DEGs) formed at oxide interfaces exhibit a unique interplay among confinement, orbital physics, spin--orbit coupling, and electronic correlations \cite{ohtomo2004high, caviglia2008electric, caviglia2010tunable, 
gariglio2018spin, barthelemy2021quasi, maiellaro2023hallmarks, guarcello2024probing, maiellaro2025theory, maiellaro2026spin, gaiardoni2026boltzmann}. Hetero-structures based on KTaO$_3$ (KTO) oxides stand out thanks to the strong spin--orbit coupling inherited from the Ta $5d$ orbitals and a remarkable tunability of their electronic properties, such as superconductivity, by electrostatic gating and interface engineering \cite{Liu2021, Hua2022_EuO_KTO110_npj, Arnault2023_SciAdv_KTO111, Mallik2023_APL_KTO111_ARPES}. The electronic structure of KTO 2DEGs is highly sensitive to the crystallographic orientation of the interface, with distinct energy and spin structures for the (001), (110), and (111) directions \cite{bruno20192d}. In particular, the (111) orientation exhibits a complex multi-orbital structure associated with the trigonal symmetry of the interface, leading to unconventional spin--orbital locking and Berry curvature effects \cite{chen2025dirac, al2025spin, zhai2023large, trama2022gate, trama2022tunable, trama2023effect,trama2021straininduced, zhao2025spin}.

One of the most striking orientation-dependent features is the superconducting critical temperature \cite{liu2023tunable, Chen2024_Orientation_LAO_KTO, cao2025geometric, maryenko2023superconductivity, ueno2011discovery, chen2025two, chen2021two, gan2023light, filippozzi2024high, cheng2025interplay, dong2026strongly, kimbell2025electricfieldcontrolsuperconducting, kim2025enhancedsuperconductivity, zhai2025nonreciprocal, cheng2025electronic, zhang2025magnetotransport, zhang2023spontaneous}. In (001)-oriented 2DEGs, superconductivity emerges only within very limited experimental conditions at high doping (see Refs.~\cite{ueno2011discovery,chen2025two}), and even then with a critical temperature $T_c$ substantially lower than in the (110) and (111) interfaces. 
The microscopic origin of this anisotropy remains open: it is unclear whether the hierarchy~\cite{liu2023tunable, Chen2024_Orientation_LAO_KTO, cao2025geometric} requires orientation-dependent pairing interactions or can arise from geometric confinement and multiorbital electronic structure alone, as suggested by the observed correlation between higher $T_c$ and stronger confinement~\cite{cheng2025interplay}.

Here we address quantitatively the impact of spatial confinement and multiorbital electronic structure on the superconducting state across all orientations. We take as our starting assumption that the pairing interaction, electronic screening, and microscopic electronic couplings are orientation-independent, and self-consistently obtain the critical temperature, accounting for a spatially-dependent order parameter through a variational technique recently developed in Ref.~\cite{trama2025self}. We correctly recover the observed hierarchy of $T_c$, showing that accounting for the inhomogeneous nature of superconductivity is crucial to determine the character of the instability.

\begin{figure}[t]
    \includegraphics[width=0.49\textwidth]{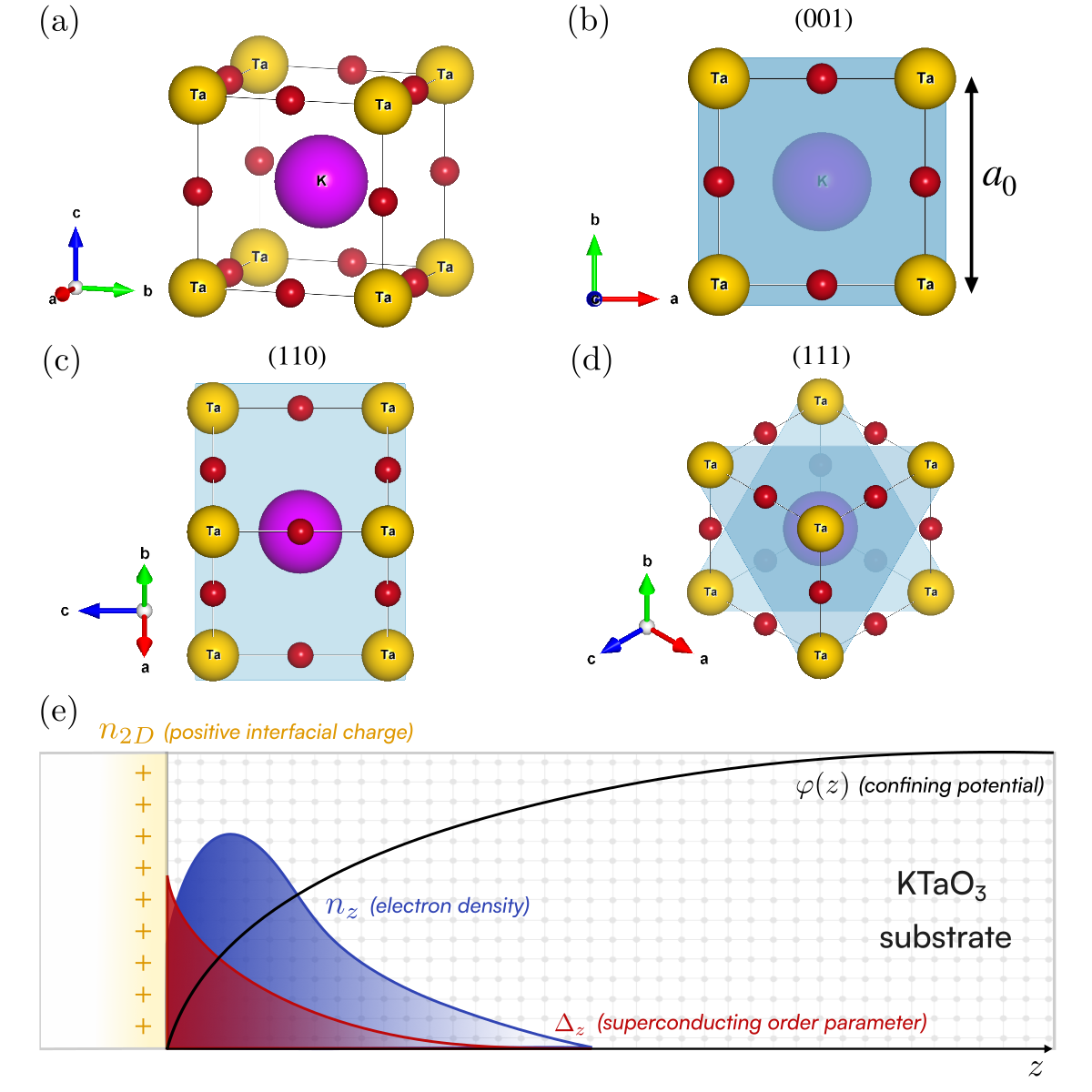}
    \caption{(a) Cubic structure of KTO. (b–d) Top view of crystalline KTO along the (001) (b), (110) (c), and (111) (d) directions. The cubic lattice constant is $a_0=3.988$ \r{A}. (e) Schematic illustration of the inhomogeneous superconductivity induced by the electronic confinement.}
    \label{fig:KTO_orientations}
\end{figure}

\begin{figure*}
    \centering
    \includegraphics[width=0.99\linewidth]{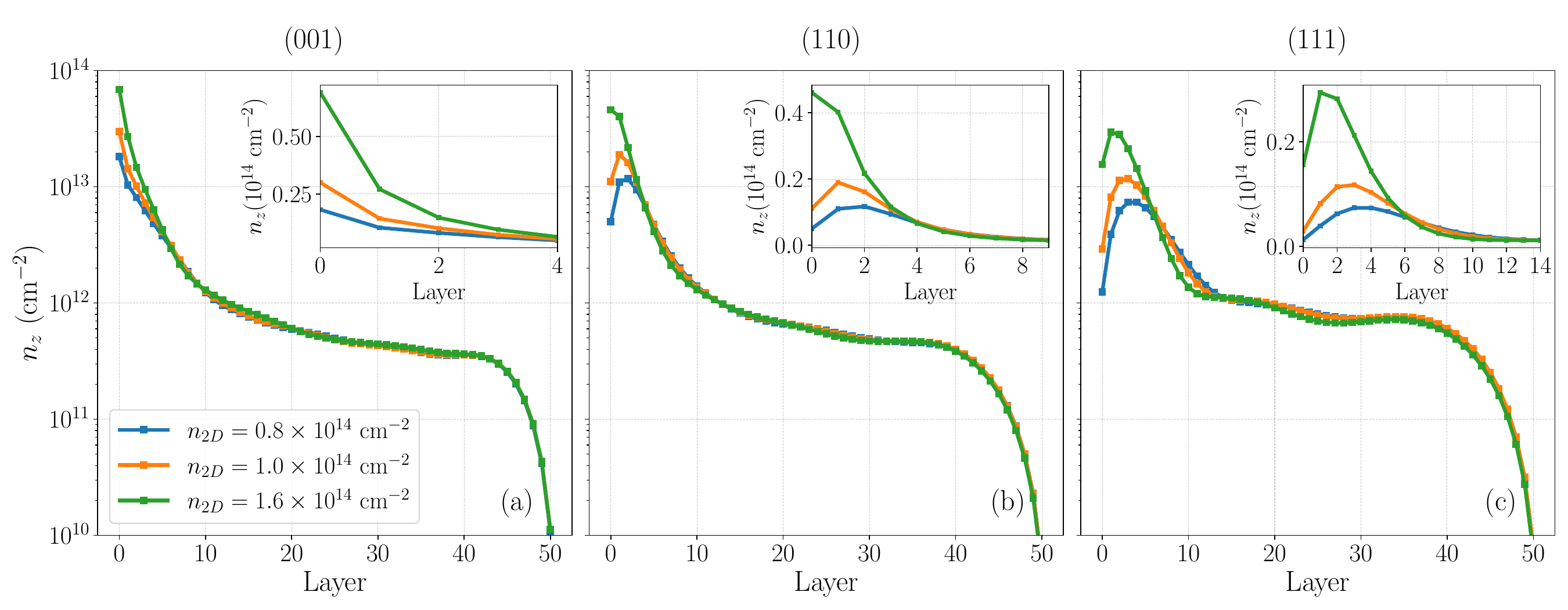}
    \caption{Electron 2D density $n_z$ for different benchmark values of $n_{2D}$ at the (001) (a), (110) (b) and (111) (c) interfaces. The insets show the details close to the interfaces.}
    \label{fig:density_evolution}
\end{figure*}

{\bf\textit{Normal electronic state.}}---The microscopic electronic interactions are taken identical across all orientations: the unified geometry, from the underlying cubic structure to the projected crystallographic cuts, is shown in Fig.~\ref{fig:KTO_orientations}.
While the (001) surface preserves the square symmetry, the (110) and (111) orientations lead to different symmetries and modified orbital overlap patterns.

\begin{figure}
    \centering
    \includegraphics[width=0.49\textwidth]{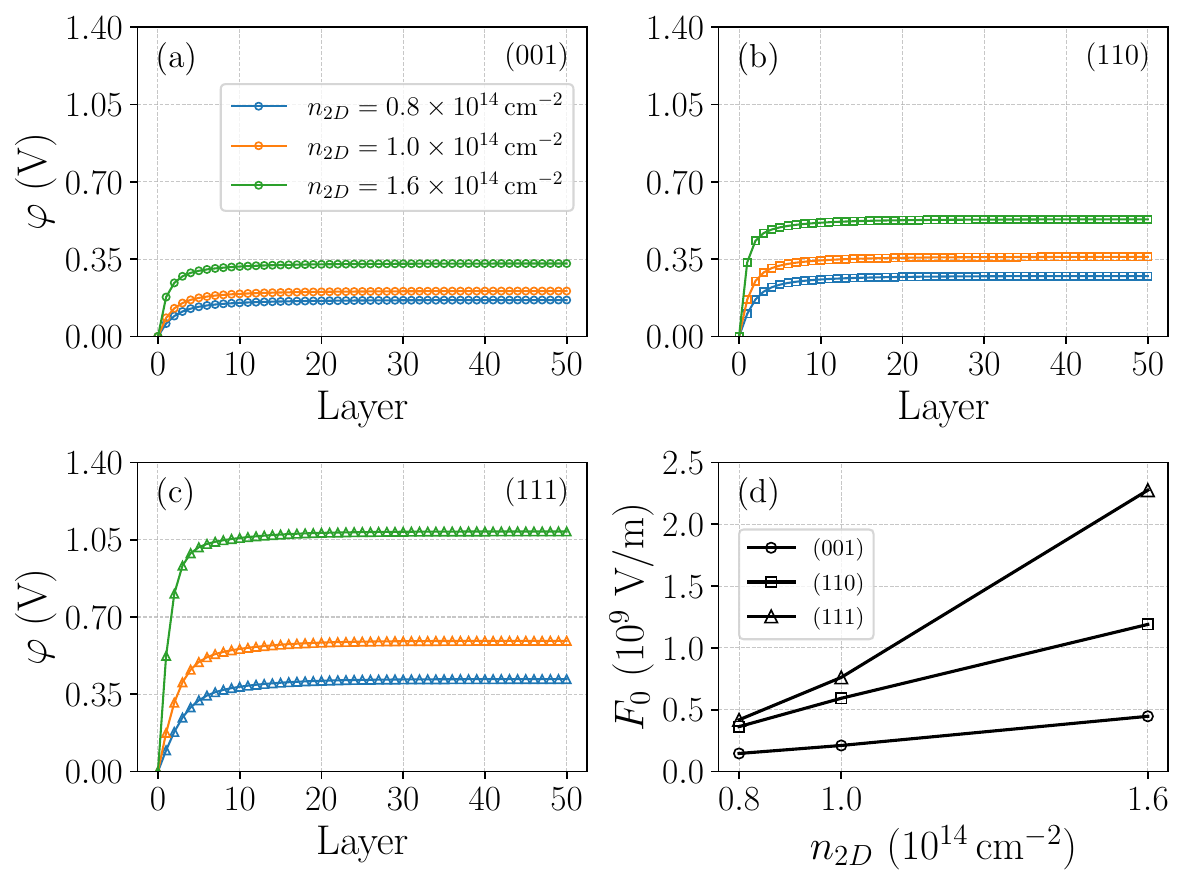}
    \caption{(a-c) Potential barrier $\varphi$ as a function of the layer position using different benchmark values of the positive charge density $n_{2D}$ for the (001) (a), (110) (b) and (111) (c) directions. (d) Interfacial electric field $F_0$ as a function of $n_{2D}$ for (001), (110) and (111) directions.}
    \label{fig:potential_evolution}
\end{figure}

The bulk electronic structure is described within a tight-binding model~\cite{trama2023effect} in terms of the conducting $t_{2g}$ orbitals denoted by $\alpha \in \{1,2,3\}=\{x,y,z\}$, corresponding to
$(d_{yz}, d_{zx}, d_{xy})$; each of the orbitals has an associated spin degree of freedom. The overall 2DEG Hamiltonian, depending on the two-dimensional quasi-momentum $\mathbf{k}$ in the plane of the 2DEG, reads
\begin{equation}
    H=H_{\rm TB}+H_{\rm SOC}+H_{\rm dist}+H_\varphi,
\end{equation}
where $H_{\rm TB}$ is obtained by projecting the same microscopic 3D bulk Hamiltonian along each crystallographic direction, $H_{\rm SOC}$ is the spin--orbit coupling term, $H_{\rm dist}$ the ${\bf k}=0$ crystal-field distortion, and $H_\varphi$ the self-consistent electrostatic confinement potential. The latter is obtained from the Poisson equation using a direction-independent dielectric function; details are given in Appendices~\ref{app:tight_binding}-\ref{app:dielectric}.

Fig.~\ref{fig:density_evolution} shows the resulting spatial profile of electron density of a slab of $N_L=51$ sites for different crystallographic orientations. 
The main control parameter is the total charge density $n_{2D}$, determining the filling of the system. 
In our model, $n_{2D}$ is represented as a fixed positive sheet charge localized at the interface, which effectively accounts for the electrostatic origin of the 2DEG, including polar discontinuity, donor defects, deposited overlayers, or top-/ionic-liquid gating~\cite{trama2023effect,ueno2011discovery,liu2013origin,Liu2021,ren2022two}.
The confinement is stronger in the (001) interface, with a stronger in-plane dispersion. Roughly, the number of layers within which the 2DEG is confined is $\Delta N_{\rm{001}}=2$ for the (001) interface, $\Delta N_{\rm{110}}=5$ for the (110) interface, and $\Delta N_{\rm{111}}=10$ for the (111) interface. 
We stress that $n_{2D}$ does not coincide with the charge density of a probed 2DEG, since it spreads along the slab.
In turn, the electrostatic potential self-consistently produced at the interface has a quite different structure among the three interfaces. Fig.~\ref{fig:potential_evolution} shows the electric potential and field for increasing density $n_{2D}$. For the same $n_{2D}$, the electric field increases from the (001) to the (110) to the (111) crystallographic cut, up to even a factor 4.
The steepest electrostatic barrier occurs along the (111) direction, in spite of the 2DEG being less spatially confined here. In this case, the stronger confinement comes rather from the anisotropic dispersion, which favors out-of-plane hopping in the (111) orientation.
\begin{figure*}
    \centering
    \includegraphics[trim={3.5cm 0 3.5cm 0}, width=0.94\textwidth]{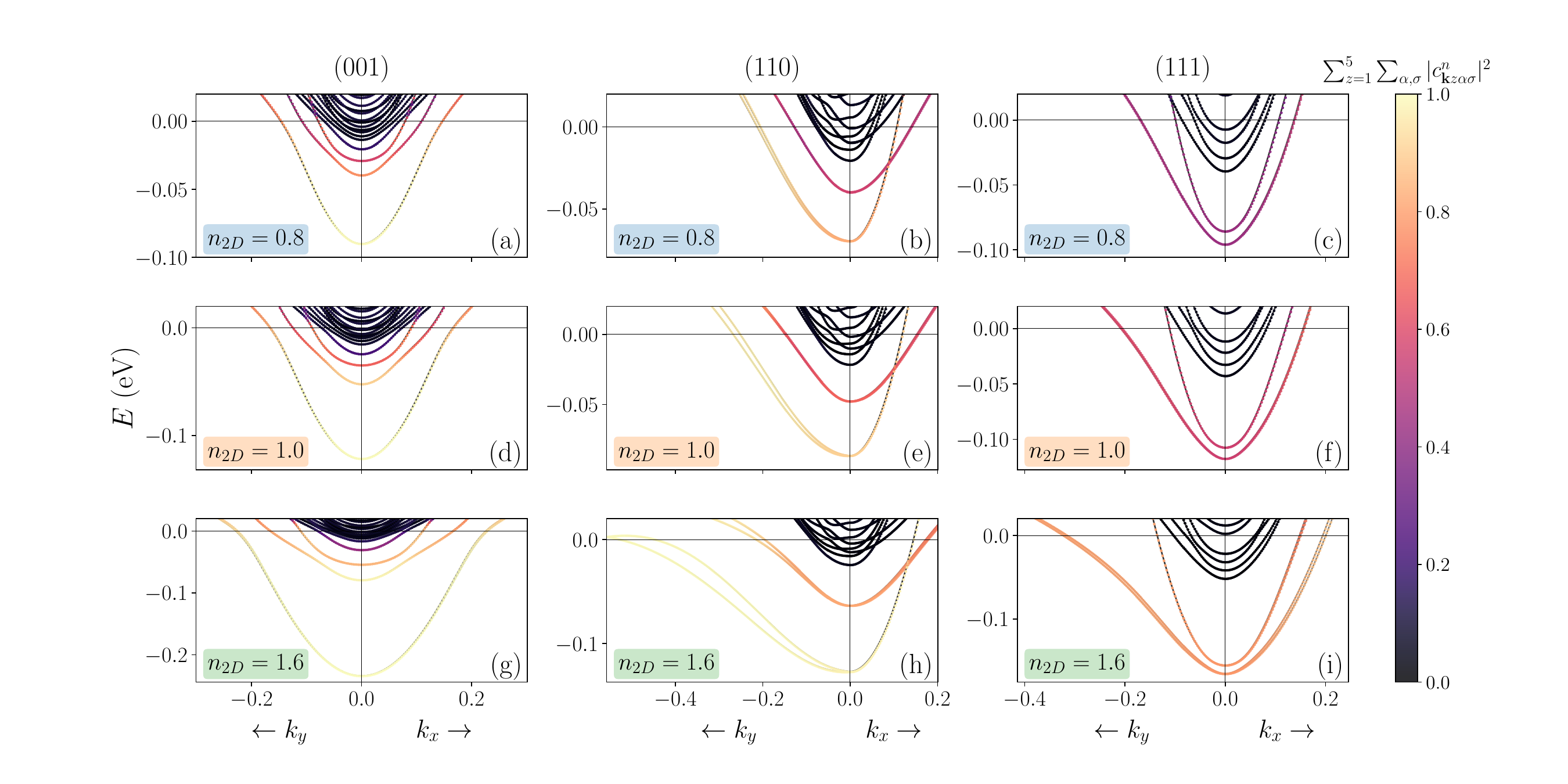}
    \caption{Layer-resolved band structure of the KTO-based heterostructure along the (a,d,g) (001), (b,e,h) (110), and (c,f,i) (111) directions for positive charge densities (a–c) $n_{2D}=0.8\times10^{14}\,\mathrm{cm}^{-2}$, (d–f) $n_{2D}=1.0\times10^{14}\,\mathrm{cm}^{-2}$, and (g–i) $n_{2D}=1.6\times10^{14} \,\mathrm{cm}^{-2}$. The color represent the probability to find a state over the first 5 layers. The quasi-momentum $k_i$ is expressed in \AA$^{-1}$. The energy zero ($E=0$) is set to the self-consistent Fermi level obtained from the electrostatic screening simulations at each $n_{2D}$.}
    \label{fig:bands_evolution}
\end{figure*}

The overall properties of the normal state are contained in the band structure, which we show in Fig.~\ref{fig:bands_evolution}. Energies are measured relative to the self-consistent chemical potential, so $E=0$ corresponds to the Fermi level at each $n_{2D}$.
The projection of the bulk $t_{2g}$ manifold onto the corresponding surface Brillouin zones produces markedly different dispersions depending on the interface orientation. For the (001) configuration, the bands exhibit a relatively stronger in-plane dispersion and a more clearly resolved confinement-induced subband structure. In contrast, the (110) and, even more prominently, the (111) orientations display a richer multi-orbital structure with enhanced band mixing and reduced effective confinement effects. Notably, even in the most localized (001) case, the chemical potential intersects bands belonging to the bulk continuum.

\begin{figure*}
    \centering
    \includegraphics[width=0.99\textwidth]{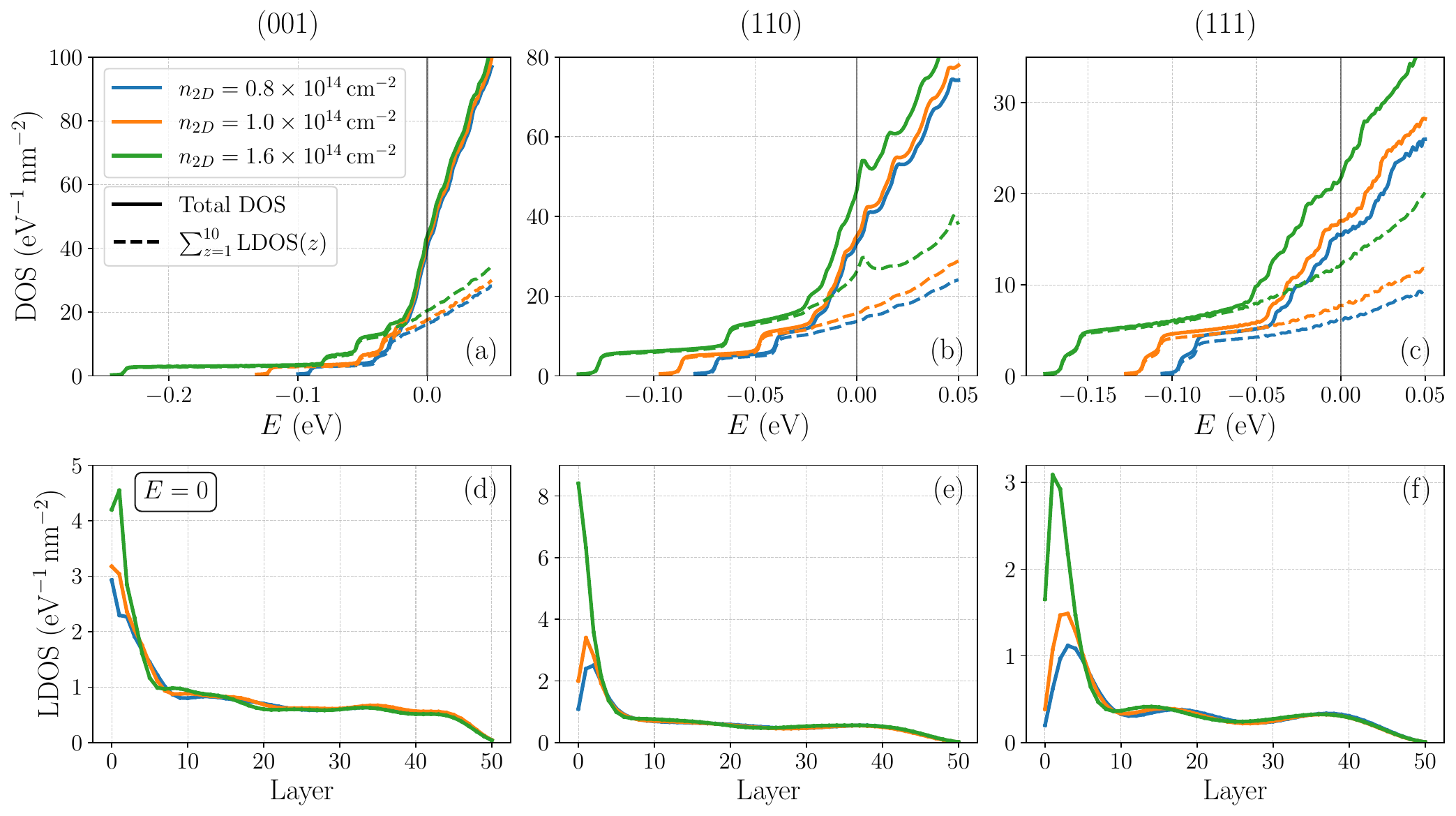}
    \caption{(a--c) Density of states (solid lines) and the sum over the first 10 layers of the layer-resolved density of states (dashed lines) as a function of energy. (d--f) Layer-resolved density of states at the self-consistent chemical potential. Panels (a,d), (b,e), and (c,f) correspond to the (001), (110), and (111) orientations, respectively.}
    \label{fig:DOS_1.6e18}
\end{figure*}

Finally, Fig.~\ref{fig:DOS_1.6e18} shows density of states (DOS) as a function of the chemical potential and the local layer-resolved density of states (LDOS) evaluated at the self-consistent chemical potential for different values of the positive interface density $n_{2D}$ and for the three crystallographic orientations. The LDOS is computed as
\begin{equation}
    {\rm LDOS}(z)=\sum_{n\,\alpha\sigma} \int \frac{d^2 \mathbf{k}}{4\pi^2} \,
    \delta(E_{n\mathbf{k}}) 
    |c^n_{\mathbf{k}z\alpha\sigma}|^2,
\end{equation}
where $E_{n\mathbf k}$ is the $n$-th band energy measured from the self-consistent chemical potential, and $c^n_{\mathbf k z\alpha\sigma}$ is the corresponding eigenvector component on layer $z$, orbital $\alpha$, and spin $\sigma$.
In practice, the Dirac delta function is approximated by a Lorentzian broadening,
$
    \pi\delta(x) \approx \tau/(x^2+\tau^2),
$
with $\tau=10^{-3}\,\mathrm{eV}$.

Once again, the LDOS shows increasing delocalization from the (001) to the (110) to the (111) configuration. 
Even though the DOS of the (001) and (110) is larger compared to the (111), the distribution of the available states along the system is displaced differently. As a benchmark, we also show the sum over the first 10 layers of the LDOS, since experimentally ARPES measurements can only probe surface properties.
This feature has a direct bearing on the superconducting instability: in a Bardeen-Cooper-Schrieffer (BCS)-like framework the superconducting gap grows with the density of states of the normal-state system. 
At the level of the bare DOS, one would therefore expect a higher $T_c$ for the (001) interface. However, the superconducting instability is spatially inhomogeneous and is controlled by a surface-weighted DOS. As shown below, this reverses the expectation and yields the highest $T_c$ for the (111) interface.

{\bf\textit{Superconducting instability.}}---Starting from a bulk three-dimensional description of a local attractive four-fermion Hamiltonian, inhomogeneous superconductivity is described within a mean-field approximation  as

\begin{equation}
H_{S}=
- U \sum_{\mathbf{k}\in\frac{1}{2}\mathrm{ BZ}}
\sum_{z\,\alpha}
\left(
\Delta_{z}\,
c^\dagger_{\mathbf{k}z \alpha \uparrow}
c^\dagger_{-\mathbf{k} z\alpha\downarrow}
+ \mathrm{h.c.}
\right)
+ 
\mathcal{A}\sum_{z}U\Delta_{z}^2,
\label{eq:mean_field_super}
\end{equation}
where $\Delta_{z}=\langle c_{iz\alpha\uparrow} c_{iz\alpha\downarrow}\rangle$. 
Here $i$ is the in-plane real-space coordinate,
$U$ is the effective pairing interaction strength, taken to be the same for all slab configurations, and $\mathcal{A}$ accounts for the different in-plane unit-cell areas associated with the various interface orientations, as we report in detail in Appendix~\ref{app:supercond}. 
We consider here a \textit{phonon-mediated pairing mechanism}, which should be regarded as a lower bound for the predicted $T_c$ within this singlet $s$-wave model. In this case, the interaction $U$ is taken to be nonzero only within an energy shell of width $\hbar \omega_i$ around the Fermi level, where $\omega_i$ is a characteristic frequency scale typical of phonon-mediated superconductivity (e.g., the Debye frequency or the frequency of soft modes~\cite{liu2023tunable, norman2026superconductivity}). 
In Appendix~\ref{app:supercond}, we also consider an \textit{instantaneous pairing mechanism}, with $U$ constant over the full energy range, which gives an upper bound for $T_c$. Since the results are qualitatively unchanged, we focus here on the standard retarded case.
In Eq.(\ref{eq:mean_field_super}), $\Delta_z$ depends explicitly on $z$, to reflect the breaking of translational invariance from the confining potential (See Fig.~\ref{fig:KTO_orientations}). To mimic the spatial profile of the electronic density in Fig.~\ref{fig:density_evolution}, we use a variational ansatz $\Delta_z=\Delta e^{-z/z_0}$ in terms of the parameters $\Delta$ and $z_0$ describing amplitude and penetration of the order parameter across the material. Different variational profiles could be introduced, but the exponential ansatz provides the minimal parametrization needed to retain the relevant physical ingredient, namely the spatial extension $z_0$ of the superconducting condensate.
To determine the self-consistent variational parameters, we minimize the free-energy functional 
\begin{dmath}
    F=
    3U\sum_{z=0}^{N_L-1}\Delta^2 e^{-2z/z_0}
    -\frac{\mathcal{P}}{\beta}
    \sum_n\int d^2\mathbf{k} 
    \left[
    \log\left(2+2\cosh\left(\beta\lambda_{n\mathbf{k}}\right)\right)
    -
    \log\left(2+2\cosh\left(\beta E_{n\mathbf{k}}\right)\right)
    \right],
    \label{eq:free_energy}
\end{dmath}
where $\beta=1/k_B T$, the factor $3$ accounts for the three orbital degrees of freedom, $n$ labels the normal-state bands, and $\mathcal P$ is geometry-dependent, with $\mathcal P_{001}=1/(4\pi^2)$, $\mathcal P_{110}=\sqrt{2}/(4\pi^2)$, and $\mathcal P_{111}=27/(8\pi^2)$ (see Appendix~\ref{app:supercond}).
The Bogoliubov--de-Gennes (BdG) eigenvalues are
\begin{equation}
\lambda_{n\mathbf{k}}=\sqrt{E_{n\mathbf{k}}^2-U^2\Delta^2 f_{n\mathbf{k}}^2(z_0)},
\label{eq:BdG_values}
\end{equation}
where $f_{n\mathbf k}(z_0)=\sum_{z\alpha\sigma}|c^n_{\mathbf k z\alpha\sigma}|^2 e^{-z/z_0}$ weights the normal-state eigenvector across the slab.

The critical interaction $U$ is obtained from the superconducting instability in the limit $\Delta\to0$, imposing
\[
\pdv[2]{F}{G}=0,
\qquad
\pdv{F}{z_0}=0,
\]
with $G=U\Delta$.

For phonon-mediated pairing mechanism
Eq.~\eqref{eq:free_energy} can be written as
\begin{dmath}
    F(z_0)\approx 3\frac{G^2}{U} \frac{1-e^{-2N_L/z_0}}{1-e^{-2/z_0}} -
    G^2\frac{\mathcal{P}}{2}\rho(z_0)
    \int_{-\omega_i}^{\omega_i}
    \frac{dE}{E}
    \tanh\left(\frac{\beta E}{2}\right),
    \label{eq:BCS_integral}
\end{dmath}
where
\begin{equation}
    \rho(z_0)=
    \sum_n\int d^2\mathbf{k} \,
    \delta(E_{n\mathbf{k}})\,
    f_{n\mathbf{k}}^2(z_0)
    \label{eq:dos_super}
\end{equation}
is the density of states at the Fermi level weighted by the factor $f^2_{n\mathbf{k}}(z_0)$. 
Thus, the instability reflects a balance between the overlap-weighted DOS and the cost of extending the condensate across the slab.

\begin{figure}
    \centering
    \includegraphics[trim={0 0 1cm 0 },width=0.49\textwidth]{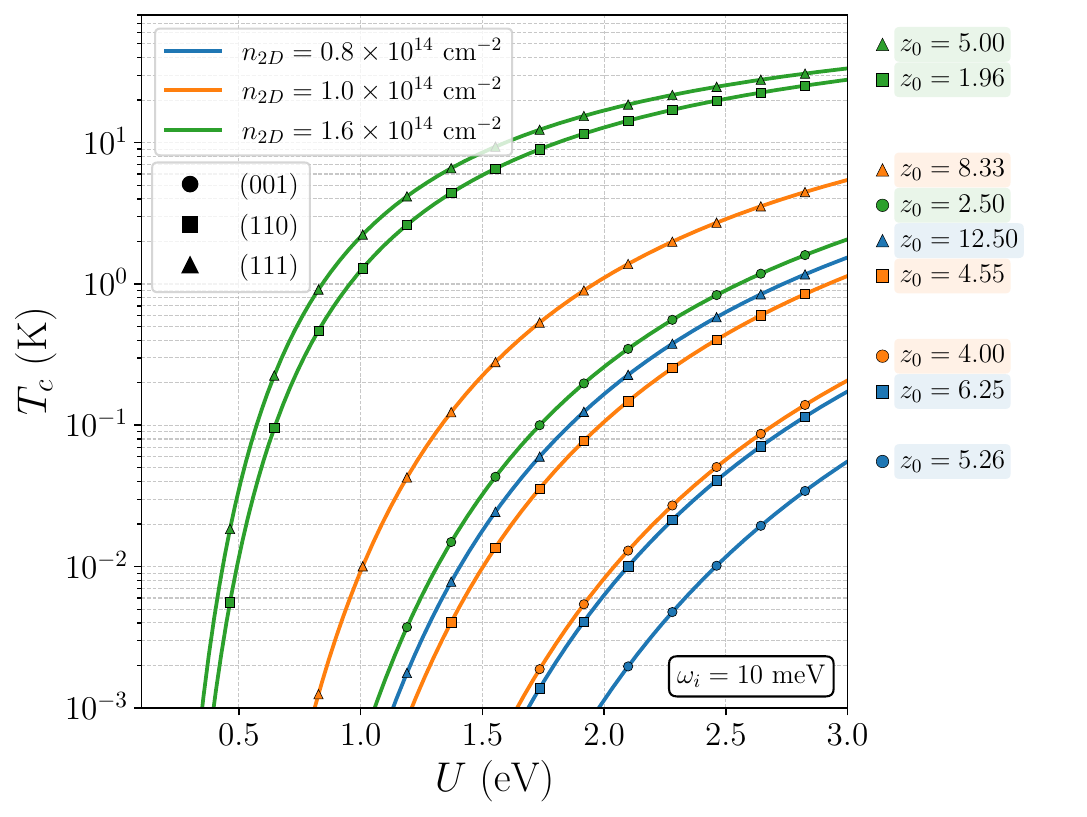}
    \caption{Critical temperature $T_c$ as a function of the interaction strength $U$ for the (001) (circles), (110) (squares), and (111) (triangles) interfaces in the phonon-mediated pairing scenario. The self-consistent penetration length $z_0$ is indicated at the end of each curve.}
    \label{fig:BCS_all.pdf}
\end{figure}

Since $U$ is fixed for all interfaces, the differences among the $T_c$ we recover across different orientations arise purely from electronic properties. 
We use $\hbar\omega_i=10$ meV as a representative phonon scale, although $\omega_i$ may also depend on the interface orientation~\cite{liu2023tunable,Chen2024_Orientation_LAO_KTO}. The resulting $T_c(U)$ is shown in Fig.~\ref{fig:BCS_all.pdf}.

At fixed electronic density, a clear hierarchy emerges among the three orientations: the (111) interface exhibits the highest $T_c$ and the largest penetration length, while the (001) orientation shows the lowest critical temperature, namely
\[
T_c^{(001)} < T_c^{(110)} < T_c^{(111)}.
\]
Moreover, there exists a range of interaction strengths for which $T_c^{(111)}\approx 1$~K, $T_c^{(110)}\approx 0.1$~K, while the (001) interface does not display a superconducting instability within accessible temperatures, consistent with experimental data in Refs.~\cite{liu2023tunable,Chen2024_Orientation_LAO_KTO}. Differences among the carrier densities in the various orientations — with the (111) interface typically hosting the largest two-dimensional electron density — may further enlarge the allowed range of $U$. Generally, increasing the electronic density leads to stronger confinement and a reduction in $z_0$. The role of inhomogeneity is indeed quite crucial: 
allowing for a spatially varying order parameter naturally captures the correct hierarchy of critical temperatures across different orientations, while also highlighting that the spatial extent of superconductivity can be sizable in these systems. In Appendix~\ref{app:details_z0} we present a representative case where the penetration depth $z_0$ is not determined self-consistently, showing how this affects the results.

{\bf\textit{Discussion.}}---
We have systematically investigated the superconducting instability arising from the electronic properties of the 2DEG at (001), (110), and (111) KTaO$_3$-based interfaces. 
To isolate the role of crystallographic orientation, we have employed a tight-binding model with identical bulk hopping amplitudes for all three geometries. 
The self-consistent confining potential has been computed for a large number of layers, using the same dielectric function fitted to experimental ARPES data~\cite{bruno20192d,zhai2023large} and considering representative values of the positive interface density $n_{2D}$.

In the normal state, the (111) orientation exhibits the steepest electrostatic barrier, even though its 2DEG is less spatially confined. This unusual behavior arises from dominant out-of-plane hopping, unlike the (001) and (110) orientations where stronger in-plane dispersion promotes interface localization. The interplay between electronic screening and anisotropic dispersion ultimately produces a stronger electric field at the (111) interface.

We have studied the onset of superconductivity using an inhomogeneous mean-field singlet $s$-wave pairing interaction. 
Allowing the superconducting order parameter to vary across the slab, we reproduce the experimentally observed hierarchy of critical temperatures, with the (111) orientation consistently displaying the highest $T_c$. 
This hierarchy is robust across two limiting pairing scenarios: a phonon-mediated (retarded) mechanism and a local instantaneous (non-retarded) interaction, which provide a lower and upper bound on the superconducting instability within our model. 
In particular, in the retarded case the results do not depend sensitively on the choice of cut-off frequency, further supporting the geometric origin of the predicted hierarchy. 
Importantly, the hierarchy of critical temperatures is not set by the interfacial electric field alone. Although the (111) interface has the strongest field, its superconducting profile is more extended across the slab. Thus, $T_c$ is controlled not simply by electrostatic confinement, but by the redistribution of low-energy spectral weight and its overlap with the layer-dependent pairing profile.

In contrast to standard hard-wall descriptions of confinement-induced superconductivity~\cite{shanenko2006size,valentinis2016rise,klimin2014interface,virtanen2019superconducting}, our mechanism relies on the smooth spatial variation of the confining potential, emphasizing the role of geometry-dependent confinement and multiorbital band structure in oxide-interface superconductivity. This is consistent with experiments on LaAlO$_3$/SrTiO$_3$, where superconductivity correlates with the occupation of confinement-induced subbands and with the shape of the confining potential, rather than with carrier density alone~\cite{smink2018correlation}.

Both the normal-state analysis and the superconducting treatment have allowed us to isolate the effects arising purely from geometric constraints and electronic structure, showing that this mechanism is sufficient to account for the qualitative hierarchy of $T_c$ without introducing orientation-dependent microscopic parameters.
In realistic systems, however, additional effects may further enhance the differences among orientations. Different surface orientations may involve distinct impurity concentrations, low-temperature lattice distortions, hopping amplitudes, dielectric responses, phonon spectra, and carrier densities, with the (111) interface typically hosting the largest 2DEG density~\cite{liu2023tunable,Chen2024_Orientation_LAO_KTO}. These effects may affect the hierarchy of $T_c$ predicted here and should be included for a fully quantitative description.
Another limitation concerns the spatial distribution of compensating charges. Here, the positive charge is assumed to be localized at the interface, whereas in real oxide interfaces donor defects and trapped electronic charges may extend into the substrate, modifying the confinement potential~\cite{biscaras2014limit,scopigno2016phase}. Including these contributions in the Poisson equation, as well as a systematic treatment of back-gating, is a natural extension of the present model and may quantitatively affect both $T_c$ and the effective superconducting thickness.
Our prediction could be further tested experimentally by extracting the effective superconducting thickness from the anisotropy of the upper critical field. In Ref.~\cite{filippozzi2024high}, a thickness $d\approx1.25$~nm was inferred for the (111) interface, consistent in order of magnitude with the characteristic extension $z_0\approx5$ layers obtained in our benchmark calculation at $n_{2D}=1.6\times10^{14}\,{\rm cm}^{-2}$.

Finally, our analysis has been restricted to singlet $s$-wave pairing. Although alternative pairing symmetries have been proposed for KTO-based systems~\cite{Arnault2023_SciAdv_KTO111, zhai2025nonreciprocal,zhang2023spontaneous}, $s$-wave pairing is expected to remain the most robust channel in the experimentally relevant dirty regime, where higher-angular-momentum components are typically suppressed.
More generally, the present framework is not tied to a specific symmetry of the order parameter. Its key ingredients—namely the multiorbital electronic structure and the finite dispersion along the $z$ direction—govern the spatial extension of the 2DEG and thus directly affect the superconducting instability, by controlling both the spatial distribution of the condensate and the structure of the pairing kernel.
The formalism can be straightforwardly generalized by considering more variational parameters and other pairing symmetries~\cite{perroni2}, allowing one to investigate how the symmetry of the order parameter modifies both its spatial profile and the resulting instability, including in the dynamic regime~\cite{perroni_din1,perroni_din2}. 
In this context, confinement plays a central role by setting the spatial structure of the electronic states participating in pairing.

\section*{Acknowledgments}
M.T. and R.C. acknowledge N. Bergeal, H. Witt, A. Caviglia, U. Filippozzi, A. K. Jaiswal, F. Romeo, C. Guarcello, A. Maiellaro, R. Capecelatro and I. Gaiardoni for fruitful discussions.
The authors acknowledge funding from IQARO (Spinorbitronic Quantum Bits in Reconfigurable 2D Oxides) project of the European Union’s Horizon Europe research and innovation programme under grant agreement n. 101115190. C.A.P. acknowledges funding from the PRIN 2022 PNRR project P2022SB73K “Superconductivity in KTaO3 Oxide-2DEG NAnodevices for Topological quantum Applications” (SONATA) financed by the European Union - Next Generation EU. R.C. acknowledges support from the PNRR MUR Project No. PE0000023-NQSTI (TOPQIN and SPUNTO)  and from the project INNOVATOR (National centre HPC, big data and quantum computing). M.T. acknowledges PNRR MUR Project No. PE0000023-NQSTI (SOC-OX CUP E63C22002180006) and computational resources from CINECA award under the ISCRA C initiative (project DISKTO HP10CCMDC7).

\appendix
\onecolumngrid

\section{Hamiltonian tight-binding model}\label{app:tight_binding}

The bulk Hamiltonian reads

\begin{dmath}
H_{\text{TB}}^{\text{Bulk}}
=
\sum_{\mathbf K}\sum_{\alpha \,\sigma}
\left[
- t_D 
\sum_{\beta}
\left(1-\delta_{\alpha\beta}\right)
\cos(K_\beta)
- t_I \cos(K_\alpha)
\right]
c^\dagger_{\mathbf K \alpha \sigma}
c^{\vphantom{\dagger}}_{\mathbf K \alpha \sigma}.
\label{eq:bulk_ham}
\end{dmath}

where $t_D$ and $t_I$ denote the nearest-neighbor (NN) and next-to-nearest-neighbor (NNN) hopping amplitudes, respectively. 
The operator $c_{\mathbf K \alpha\sigma}$ annihilates an electron with crystal momentum $\mathbf K$, orbital $\alpha$, and spin $\sigma$.
The momentum $\mathbf K$ is expressed in Cartesian coordinates, with components $K_\alpha$ measured in units of $1/a_0$, where $a_0 = 3.988$~\r{A} is the cubic lattice constant of KTO.
For each crystallographic direction, we rotate the coordinate system and discretize the Hamiltonian along the direction transverse to the interface~\cite{trama2023effect}, according to the structures shown in Fig.~\ref{fig:KTO_orientations}. For each direction, we denote by $\mathbf{k}=(k_x,k_y)$ its quasi-momentum (in units of the inverse of in-plane lattice constants) and by $z$ the corresponding transverse direction. 

We include the atomic spin-orbit coupling $H_{\rm SOC}$ in the form
\begin{equation}
    H_{\rm SOC}= \frac{\lambda}{2}
\sum_{\mathbf{k}}
\sum_{z\,\alpha\beta\gamma\,\sigma \sigma'}
i \, \varepsilon_{\alpha\beta\gamma }
\, c^{\dagger}_{\mathbf{k} z\alpha \sigma}
\, \sigma^{\gamma}_{\sigma \sigma'}
\, c^{\vphantom{\dagger}}_{\mathbf{k} z\beta \sigma'},
\label{eq:soc}
\end{equation}
where $\lambda\approx0.27$~eV~\cite{zhai2023large}.

For each slab orientation, we also consider a $\mathbf{k}=0$ perturbation $H_{\rm dist}$ describing crystal-field distortions:
\begin{equation}
    H^{(001)/(110)}_{\rm dist}= \frac{\delta}{2}\sum_{\mathbf{k}} \sum_{ z \,\sigma}
c^{\dagger}_{\mathbf{k} z3\sigma}
\, c_{\mathbf{k}z3\sigma},
\label{eq:tetragonal}
\end{equation}
\begin{equation}
    H^{(111)}_{\rm dist}= \frac{\delta}{2}\sum_{\mathbf{k}} \sum_{z\, \alpha \neq \beta\,\sigma}
c^{\dagger}_{\mathbf{k} z\alpha\sigma}
\, c^{\vphantom{\dagger}}_{\mathbf{k} z\beta\sigma},
\label{eq:trigonal}
\end{equation}
corresponding to tetragonal and trigonal crystal fields, respectively. We set $\delta=-10$~meV, a typical magnitude for such distortions~\cite{trama2021straininduced}.  
For the (001) and (110) orientations this distortion is typically negligible, since the $d_{xy}$ orbital is already split from $d_{yz}$ and $d_{zx}$. In contrast, for the (111) orientation the trigonal field lifts the orbital degeneracy, leading to qualitative changes in the low-filling band structure.

Finally, we introduce a $z$-dependent electrostatic potential describing confinement at the interface
\begin{equation}
    H_{\varphi}=\sum_{\mathbf{k}}\sum_{z\,\alpha\,\sigma} 
    \varphi(z)\,
    c^{\dagger}_{\mathbf{k} z\alpha\sigma}
    c^{\vphantom{\dagger}}_{\mathbf{k} z\alpha\sigma},
\label{eq:ele_pot}
\end{equation}
where $\varphi(z)$ is determined self-consistently.
The term in Eq.~\eqref{eq:ele_pot} breaks inversion symmetry, leading to the emergence of the orbital Rashba effect~\cite{trama2022gate,trama2022tunable,zhai2023large}. Its amplitude is proportional to the electric field at each layer and couples different layers. In Section~\ref{app:rahsba} we describe both NN and NNN Rashba hoppings arising along all three crystallographic directions. Unlike simplified monolayer or four-band models commonly used for (001) and (111) interfaces, we include all couplings up to NNN.

In order to ensure a consistent comparison among interfaces, we adopt a common set of benchmark parameters. We fix the hopping amplitudes $t_D$ and $t_I$ of 
Eq.~\eqref{eq:bulk_ham} using the (111) slab as a reference. From ARPES data~\cite{bruno20192d} combined with an effective electrostatic potential model~\cite{zhai2023large}, we extract $t_D = 0.78$~eV and $t_I = 0.029$~eV.
The self-consistent electric potential is obtained through the iterative scheme described in Ref.~\cite{trama2023effect}. The procedure couples the electronic density $n_z$, computed from the squared modulus of the layer-resolved wave functions after diagonalization to the discrete Maxwell equations along the finite direction:
\begin{equation}
\begin{cases}
\displaystyle
\varphi(z) = - a^\prime \sum_{l=1}^{z} F_{l}, \\[6pt]
\varepsilon_0 \, \varepsilon(F_z) F_z = D_z, \\[6pt]
D_z = |e|\left(n_{2D} - \sum_{l=1}^{z} n_l \right),
\end{cases}
\label{eq:maxwell}
\end{equation}
where $a^\prime$ is the interlayer spacing along the $z$ direction, $F_z$ is the electric field, and $n_{2D}$ the positive interface charge density, treated as a free parameter and screened by the electronic charge in the slab.
The choice of $\varepsilon(F_z)$ is discussed in Section~\ref{app:dielectric}.

\section{Orbital Rashba terms for different orientations}\label{app:rahsba}

We derive the matrices describing the orbital Rashba coupling
arising from inversion-symmetry breaking along the (001), (110), and (111)
crystallographic directions.

In simplified 2D treatments,
the orbital Rashba term is often written, for the (001) interface, in the form
\begin{equation}
    h^{\rm R}\propto
    \begin{pmatrix}
    0 & 0 & \sin(k_x)\\
    0 & 0 & \sin(k_y)\\
    -\sin(k_x) & -\sin(k_y) & 0
    \end{pmatrix},
\end{equation}
in the $t_{2g}=\{d_{yz},d_{zx},d_{xy}\}$ basis.
However, this expression captures only the lowest-order in-plane processes and
is not sufficient for a multi-layer slab geometry.

Following the approach of Ref.~\cite{shanavas2014theoretical} and its
multi-layer generalization in Ref.~\cite{trama2022gate}, we derive the full
Rashba matrices including both NN and NNN processes. 
For a given orientation of the interfacial
electric field, we first determine the linear-in-field correction to the local
$t_{2g}$ orbital wavefunctions. We then evaluate the corresponding matrix
elements of the unperturbed tight-binding Hamiltonian in the three-dimensional
Brillouin zone, obtaining an effective orbital Rashba Hamiltonian
$H^{\rm R}(\mathbf{K})$.
To construct the slab model, we split $\mathbf{K}=(\mathbf{k},k_z)$ into
in-plane momentum $\mathbf{k}$ and the component $k_z$ orthogonal to the
interface, and discretize the perpendicular direction. In this step, phase
factors $e^{\pm ik_z}$ are identified with interlayer hoppings, allowing us
to decompose $H^{\rm R}$ into in-plane and out-of-plane (interlayer) blocks.

\subsubsection{(001) direction}

For an electric field oriented along $\hat{z}$, the linear correction to the
$t_{2g}$ orbitals reads~\cite{shanavas2014theoretical}
\begin{equation}
\ket{d_{yz}'} = \frac{\eta_p}{\sqrt{5}}\ket{p_y}, \quad
\ket{d_{zx}'} = \frac{\eta_p}{\sqrt{5}}\ket{p_x}, \quad
\ket{d_{xy}'} = 0
\label{eq:d001}
\end{equation}
Here $\eta_p$ contains the radial matrix element induced by the electric field and its dependency by the electric potential $\varphi(z)$ as $\eta_p\sim\frac{\varphi(z)}{a^\prime\, 10~\mathrm{eV/nm}}$, where $a^\prime$ being the distance between two adjacent layers.

Let $H_0$ denote the unperturbed Hamiltonian, i.e. the sum of
Eqs.~\eqref{eq:bulk_ham}–\eqref{eq:ele_pot}. To linear order in the
field-induced correction of the orbital wavefunctions, the matrix elements in
the $\{d_\alpha(\mathbf{K})\}$ basis are

\begin{equation}
\mel{d_{\alpha}(\mathbf{K})}{H_0}{d_{\beta}(\mathbf{K})}
=
\sum_{\mathbf{R},s={x,y,z}}
\left(
c_{\alpha,s}^*
\mel{p_s}{H_0}{d_\beta(-\mathbf{R})}
+
c_{\beta,s}
\mel{p_s}{H_0}{d_\alpha(-\mathbf{R})}^*
\right)
e^{i\mathbf{K}\cdot\mathbf{R}},
\label{eq:rashbamatrix}
\end{equation}
where $\alpha,\beta$ running over the orbital degrees of freedom and $c_{\alpha,s}$ are the coefficients defined in
Eqs.~\eqref{eq:d001}.

For convenience, we introduce the Slater–Koster combinations
\begin{equation}
p_1 \equiv V_{pd\pi},
\qquad
p_3 \equiv -2V_{pd\pi}+\sqrt{3}\,V_{pd\sigma},
\label{eq:p1p3_def}
\end{equation}
where $V_{pd\pi}$ and $V_{pd\sigma}$ are the Slater--Koster $pd\pi$ and $pd\sigma$
overlap integrals~\cite{slater1954simplified}.

\paragraph{Nearest neighbors.}

Restricting to NN $\mathbf{R}=\pm a_0\hat{\xi}$, with
$\hat{\xi}\in\{\hat{x},\hat{y},\hat{z}\}$, we obtain

\begin{equation}
H_a^{\rm R}=
p_1\eta_{p}(\sqrt{2})^{7/2}\frac{2i}{\sqrt{5}}\\
\begin{pmatrix}
0 & 0 & \sin(K_x)\\
0 & 0 & \sin(K_y)\\
-\sin(K_x) & -\sin(K_y) & 0
\end{pmatrix},
\label{eq:stand_rashba}
\end{equation}
where $\mathbf{K}$ is measured in units of $a_0^{-1}$. 
The factor $(\sqrt{2})^{7/2}$ accounts for the relative scaling between
NN and NNN contributions.
Equation~\eqref{eq:stand_rashba} corresponds to the standard orbital Rashba
structure for the (001) orientation.

\paragraph{Next-to-nearest neighbors.}

For NNN
$\mathbf{R}=\pm a_0(\hat{\xi}+\lambda\hat{\mu})$,
with $\lambda=\pm1$ and $\hat{\mu}={\hat{x},\hat{y},\hat{z}}$
with
$\hat{\mu}\neq\hat{\xi}$, the matrix is 
\begin{equation}
H_{\sqrt{2}a}^{\rm R}
=
\eta_p \frac{4i}{\sqrt{5}}
\begin{pmatrix}
0 & 0 & F_x(\mathbf{K}) \\
0 & 0 & F_y(\mathbf{K}) \\
- F_x(\mathbf{K}) & - F_y(\mathbf{K}) & 0
\end{pmatrix}.
\end{equation}
having defined
\begin{align}
F_x(\mathbf{K}) &= \left[(p_1+p_3)\cos(K_y) + p_1\cos(K_z)\right]\sin(K_x),\\
F_y(\mathbf{K}) &= \left[(p_1+p_3)\cos(K_x) + p_1\cos(K_z)\right]\sin(K_y),
\end{align}

\paragraph{Slab decomposition.}

We now decompose $\mathbf{K}=(\mathbf{k},K_z)$, where
$\mathbf{k}=(k_x,k_y)$ is the in-plane momentum (here the redefinition is purely formally in order to unify the notation and remember that we switched to a 2D layered system).
Upon discretization of the $z$ direction,
$e^{\pm iK_z}$ are identified with interlayer hoppings.

Since $H_a^{\rm R}$ does not depend on $K_z$, it contributes only to
in-plane processes:

\begin{equation}
H^{\rm R}_{a,\mathrm{in\text{-}plane}}(\mathbf{k})
=
p_1\eta_{p}(\sqrt{2})^{7/2}\frac{2i}{\sqrt{5}}\\
\begin{pmatrix}
0 & 0 & \sin(k_x)\\
0 & 0 & \sin(k_y)\\
-\sin(k_x) & -\sin(k_y) & 0
\end{pmatrix}.
\label{eq:001_inplane_NN}
\end{equation}

For the NNN contribution, using
$\cos K_z=\tfrac{1}{2}(e^{iK_z}+e^{-iK_z})$,
we separate in-plane and interlayer terms as

\begin{equation}
H_{\sqrt{2}a}^{\rm R}(\mathbf{K})
=
H^{\rm R}_{\sqrt{2}a,\mathrm{in\text{-}plane}}(\mathbf{k})
+
e^{iK_z}H^{\rm R}_{\sqrt{2}a,\mathrm{out\text{-}of\text{-}plane}}(\mathbf{k})
+
e^{-iK_z}
\left[H^{\rm R}_{\sqrt{2}a,\mathrm{out\text{-}of\text{-}plane}}(\mathbf{k})\right]^{\!\dagger}.
\label{eq:001_slab_decomposition}
\end{equation}

Dropping the $\cos(K_z)$ term yields the purely in-plane NNN matrix:

\begin{equation}
\tilde F_x(\mathbf{k}) = (p_1+p_3)\cos(k_y)\sin(k_x),\quad
\tilde F_y(\mathbf{k}) = (p_1+p_3)\cos(k_x)\sin(k_y),
\end{equation}

\begin{equation}
H^{\rm R}_{\sqrt{2}a,\mathrm{in\text{-}plane}}(\mathbf{k})
=
\eta_p \frac{4i}{\sqrt{5}}
\begin{pmatrix}
0 & 0 & \tilde F_x(\mathbf{k}) \\
0 & 0 & \tilde F_y(\mathbf{k}) \\
-\tilde F_x(\mathbf{k}) & -\tilde F_y(\mathbf{k}) & 0
\end{pmatrix}.
\label{eq:001_inplane_NNN}
\end{equation}

The remaining $\cos(K_z)$ contribution generates the interlayer hopping:

\begin{equation}
H^{\rm R}_{\sqrt{2}a,\mathrm{out\text{-}of\text{-}plane}}(\mathbf{k})
=
\eta_p \frac{4i}{\sqrt{5}}\frac{p_1}{2}\\
\begin{pmatrix}
0 & 0 & \sin(k_x)\\
0 & 0 & \sin(k_y)\\
-\sin(k_x) & -\sin(k_y) & 0
\end{pmatrix}.
\label{eq:001_outofplane_NNN}
\end{equation}

\subsubsection{(110) direction}

For an electric field oriented along $(\hat{x}+\hat{y})/\sqrt{2}$, the linear
corrections to the $t_{2g}$ orbitals read
\begin{equation}
    \ket{d_{yz}'} = \frac{\eta_p}{\sqrt{10}}\ket{p_z}, \quad
    \ket{d_{zx}'} = \frac{\eta_p}{\sqrt{10}}\ket{p_z}, \quad
    \ket{d_{xy}'} = \frac{\eta_p}{\sqrt{10}}\left(\ket{p_x}+\ket{p_y}\right).
\end{equation}

We first evaluate the Rashba matrix elements in the cubic coordinate system
$(x,y,z)$ using Eq.~\eqref{eq:rashbamatrix}.

\paragraph{Nearest neighbors.}

For $\mathbf{R}=\pm a_0\hat{\xi}$, we define
\begin{equation}
A(\mathbf{K}) = p_1 \big(\sin K_x - \sin K_y\big),\quad
B(\mathbf{K}) = p_1 \sin K_z,
\end{equation}
and obtain
\begin{equation}
H_a^{\rm R}
=
\frac{2i}{\sqrt{10}}(\sqrt{2})^{7/2}
\begin{pmatrix}
0 & A(\mathbf{K}) & -B(\mathbf{K}) \\
- A(\mathbf{K}) & 0 & -B(\mathbf{K}) \\
B(\mathbf{K}) & B(\mathbf{K}) & 0
\end{pmatrix}.
\end{equation}

\paragraph{Next-to-nearest neighbors.}

For $\mathbf{R}=\pm a_0(\hat{\xi}+\lambda\hat{\mu})$,
with $\lambda=\pm1$ and $\hat{\mu}\neq\hat{\xi}$, we introduce

\begin{align}
C(\mathbf{K}) &=
2 \Big[
p_1 \cos\!\left(\tfrac{K_x-K_y}{2}\right)
+ (p_1+p_3)\cos\!\left(\tfrac{K_x+K_y}{2}\right)\cos K_z
\Big]
\sin\!\left(\tfrac{K_x-K_y}{2}\right),\\[4pt]
D(\mathbf{K}) &=
\Big[ p_1 \cos K_x + (p_1+p_3)\cos K_y \Big]\sin K_z,\\
\bar{D}(\mathbf{K}) &=
\Big[ (p_1+p_3)\cos K_x + p_1 \cos K_y \Big]\sin K_z.
\end{align}

The corresponding matrix reads
\begin{equation}
H_{\sqrt{2}a}^{\rm R}(\mathbf{K})
=
\frac{4i}{\sqrt{10}}
\begin{pmatrix}
0 & C(\mathbf{K}) & -D(\mathbf{K}) \\
- C(\mathbf{K}) & 0 & -\bar D(\mathbf{K}) \\
D(\mathbf{K}) & \bar D(\mathbf{K}) & 0
\end{pmatrix}.
\end{equation}

\paragraph{Rotation to the (110) slab coordinates.}

We now express the result in the orthonormal basis
\begin{equation}
\hat{\mathbf{e}}_{1}=\frac{1}{\sqrt{2}}(1,-1,0),\qquad
\hat{\mathbf{e}}_{2}=(0,0,1),\qquad
\hat{\mathbf{e}}_{3}=\frac{1}{\sqrt{2}}(1,1,0),
\end{equation}
where $\hat{\mathbf{e}}_{3}$ is normal to the interface.

The crystal momentum transforms as
\begin{equation}
K_x = \frac{k_z + k_y}{\sqrt{2}},\qquad
K_y = \frac{k_z - k_y}{\sqrt{2}},\qquad
K_z = k_x,
\label{eq:110_momentum_map}
\end{equation}
where $\mathbf{k}=(k_x,k_y)$ is the in-plane momentum of the slab.

Upon discretization of the perpendicular direction, the factors
$e^{\pm i k_z/\sqrt{2}}$ correspond to hoppings between adjacent layers.

\paragraph{Slab decomposition.}

The NN contribution does not generate interlayer hoppings and therefore
enters only the in-plane block.

The NNN term produces both in-plane and interlayer processes, which we
decompose as
\begin{equation}
H^{\rm R}_{\sqrt{2}a}(\mathbf{K})
=
H^{\rm R}_{\sqrt{2}a,\mathrm{in\text{-}plane}}(\mathbf{k})
+
e^{ik_z/\sqrt{2}}
H^{\rm R}_{\sqrt{2}a,\mathrm{out\text{-}of\text{-}plane}}(\mathbf{k})
+
e^{-ik_z/\sqrt{2}}
\left[
H^{\rm R}_{\sqrt{2}a,\mathrm{out\text{-}of\text{-}plane}}(\mathbf{k})
\right]^{\!\dagger}.
\end{equation}

The total in-plane Rashba block is
\begin{equation}
H^{\rm R}_{\mathrm{in\text{-}plane}}(\mathbf{k})
=
\eta_p p_1\frac{2i}{\sqrt{10}}
\begin{pmatrix}
0 & a(\mathbf{k}) & -b(\mathbf{k})\\
-a(\mathbf{k}) & 0 & -b(\mathbf{k})\\
b(\mathbf{k}) & b(\mathbf{k}) & 0
\end{pmatrix},
\end{equation}
where
\begin{equation}
a(\mathbf{k}) = 2\sin\!\left(\sqrt{2}\,k_y\right),\quad
b(\mathbf{k}) = (\sqrt{2})^{7/2}\sin(k_x).
\end{equation}

The interlayer Rashba hopping reads
\begin{dmath}
H^{\rm R}_{\sqrt{2}a,\mathrm{out\text{-}of\text{-}plane}}(\mathbf{k}) = \frac{2i}{\sqrt{10}}
\begin{pmatrix}
0 & \sin\!\left(\frac{k_y}{\sqrt{2}}\right)u(\mathbf{k}) & -\sin(k_x)\,v(\mathbf{k})\\
-\sin\!\left(\frac{k_y}{\sqrt{2}}\right)u(\mathbf{k}) & 0 & -\sin(k_x)\,v^*(\mathbf{k})\\
\sin(k_x)\,v(\mathbf{k}) & \sin(k_x)\,v^*(\mathbf{k}) & 0
\end{pmatrix},
\end{dmath}
with
\begin{align}
u(\mathbf{k})
&=
(\sqrt{2})^{7/2} p_1 
+ 2(p_1+p_3)
\cos(k_x),\\[6pt]
v(\mathbf{k})
&=
(2p_1+p_3)
\cos\!\left(\frac{k_y}{\sqrt{2}}\right)
- i\,p_3
\sin\!\left(\frac{k_y}{\sqrt{2}}\right).
\end{align}

\subsubsection{(111) direction}

For an electric field oriented along the $(111)$ direction, the linear
corrections to the $t_{2g}$ orbitals read
\begin{equation}
    \ket{d_{xy}'} = \frac{\eta_p}{\sqrt{15}}\left(\ket{p_x}+\ket{p_y}\right),\quad
    \ket{d_{yz}'} = \frac{\eta_p}{\sqrt{15}}\left(\ket{p_y}+\ket{p_z}\right),\quad
    \ket{d_{zx}'} = \frac{\eta_p}{\sqrt{15}}\left(\ket{p_z}+\ket{p_x}\right).
\end{equation}

The Rashba matrices for the (111) orientation were previously derived
for a bilayer geometry in Refs.~\cite{trama2022gate,trama2022tunable},
using a different gauge convention.
Here we adopt a gauge consistent with the present slab formulation
and extend the construction to an arbitrary number of layers.
The matrices reported below are therefore obtained by an appropriate
regauging of those in Refs.~\cite{trama2022gate,trama2022tunable}
and by generalizing the bilayer construction to the multilayer case,
including NNN processes coupling layers
separated by two lattice spacings.

\paragraph{Coordinate system.}

We introduce the orthonormal basis
\begin{equation}
\hat{\mathbf e}_1=\dfrac{1}{\sqrt{2}}(1,-1,0), \quad
\hat{\mathbf e}_2=\dfrac{1}{\sqrt{6}}(1,1,-2), \quad
\hat{\mathbf e}_3=\dfrac{1}{\sqrt{3}}(1,1,1)
\end{equation}

where $\hat{\mathbf e}_3$ is normal to the interface and
$(\hat{\mathbf e}_1,\hat{\mathbf e}_2)$ span the (111) plane.

The lattice vector decomposes as
$\mathbf R = u_1 \hat{\mathbf e}_1 + u_2 \hat{\mathbf e}_2 + u_3 \hat{\mathbf e}_3$; 
the in-plane Ta–Ta distance is
$
\tilde a = \sqrt{\frac{2}{3}}\,a_0,
$
while the interlayer spacing along $\hat{\mathbf e}_3$ is
$
\tilde a' = \frac{a_0}{\sqrt3}
$.

In the following, the in-plane momentum
$\mathbf k=(k_x,k_y)$ denotes the components along
$(\hat{\mathbf e}_1,\hat{\mathbf e}_2)$
and is expressed in units of $\tilde a^{-1}$.
The perpendicular component $k_z$ is measured in units of $(\tilde a')^{-1}$.

\paragraph{In-plane Rashba coupling.}

Defining
\begin{equation}
    s_i(\mathbf{k})=\sin\kappa_i(\mathbf{k})
\end{equation}
with
\begin{equation}
\kappa_1=-\dfrac{\sqrt{3}}{2}k_x+\dfrac{3}{2}k_y, \quad
\kappa_2=-\dfrac{\sqrt{3}}{2}k_x-\dfrac{3}{2}k_y, \quad
\kappa_3=\sqrt{3}\,k_x
\end{equation}
the intra-layer Rashba Hamiltonian takes the form
\begin{equation}
H^{\rm R}_{\sqrt{2}a,\mathrm{in\text{-}plane}}(\mathbf{k})
=
H_{\pi}(\mathbf{k})+H_{\sigma}(\mathbf{k}),
\end{equation}
with
\begin{equation}
H_{\pi}(\mathbf{k})
=
\eta_p\frac{2i}{\sqrt{15}}p_1
\begin{pmatrix}
0 & \alpha_{12} & \alpha_{13}\\
-\alpha_{12} & 0 & \alpha_{23}\\
-\alpha_{13} & -\alpha_{23} & 0
\end{pmatrix},
\end{equation}
\begin{dmath}
H_{\sigma}(\mathbf{k})
=
\eta_p\frac{2i}{\sqrt{15}}(2p_1+p_3)
\begin{pmatrix}
0 & \beta_{12} & \beta_{13}\\
-\beta_{12} & 0 & \beta_{23}\\
-\beta_{13} & -\beta_{23} & 0
\end{pmatrix},
\end{dmath}
where
\begin{align}
\alpha_{12} &= -2s_1+s_2+s_3, &
\beta_{12} &= -s_2-s_3, \\
\alpha_{13} &= -2s_3+s_1-s_2, &
\beta_{13} &= -s_1+s_2, \\
\alpha_{23} &= 2s_2-s_1+s_3, &
\beta_{23} &= s_1-s_3.
\end{align}

In contrast to the (001) and (110) cases, the purely in-plane Rashba
coupling for the (111) geometry originates entirely from NNN processes.

\paragraph{Slab decomposition.}

In the slab geometry, the full Rashba Hamiltonian can be written as
    \begin{equation}
H^{\rm R}(\mathbf{K})
=
H^{\rm R}_{\sqrt{2}a,\mathrm{in\text{-}plane}}(\mathbf{k})
+
e^{ik_z}
H^{\rm R}_{a,\mathrm{out\text{-}of\text{-}plane}}(\mathbf{k})
+
e^{-ik_z}
H^{\rm R\dagger}_{a,\mathrm{out\text{-}of\text{-}plane}}(\mathbf{k})
+
e^{i2k_z}
H^{\rm R}_{\sqrt{2}a,\mathrm{out\text{-}of\text{-}plane}}(\mathbf{k})
+
e^{-i2k_z}
H^{\rm R\dagger}_{\sqrt{2}a,\mathrm{out\text{-}of\text{-}plane}}(\mathbf{k}).
    \end{equation}

Here the NN term connects adjacent layers, while the NNN term
couples layers separated by two lattice spacings.

\paragraph{Interlayer hopping (nearest neighbors).} 
In the present gauge, the corresponding matrix reads

\begin{dmath}
H^{\rm R}_{a,\mathrm{out\text{-}of\text{-}plane}}(\mathbf{k})
=
\eta_p\,\frac{p_1}{\sqrt{15}}(\sqrt{2})^{7/2}
\begin{pmatrix}
0 & h_{12}(\mathbf{k}) & h_{13}(\mathbf{k})\\
-h_{12}(\mathbf{k}) & 0 & h_{23}(\mathbf{k})\\
-h_{13}(\mathbf{k}) & -h_{23}(\mathbf{k}) & 0
\end{pmatrix},
\end{dmath}
with
\begin{align}
h_{12}(\mathbf{k})
&=
-2i\,e^{-\frac{i}{2}k_y}\,
\sin\!\left(\frac{\sqrt{3}}{2}k_x\right),\\
h_{13}(\mathbf{k})
&=
-e^{ik_y}
+
e^{-\frac{i}{2}(\sqrt{3}k_x+k_y)},\\
h_{23}(\mathbf{k})
&=
-e^{ik_y}
+
e^{\frac{i}{2}(\sqrt{3}k_x-k_y)}.
\end{align}

\paragraph{Interlayer hopping (next-to-nearest neighbors).}
Because of the triangular stacking of the (111) planes,
NNN Slater–Koster processes generate additional
Rashba terms connecting layers separated by two lattice spacings.
These contributions are essential in a multilayer geometry,
where they provide the leading correction beyond the bilayer model.
The corresponding matrix is

\begin{dmath}
H^{\rm R}_{\sqrt{2}a,\mathrm{out\text{-}of\text{-}plane}}(\mathbf{k})
=
\frac{\eta_p}{\sqrt{15}}\,(p_1+p_3)
\begin{pmatrix}
0 & g_{12}(\mathbf{k}) & g_{13}(\mathbf{k})\\
-g_{12}(\mathbf{k}) & 0 & g_{23}(\mathbf{k})\\
-g_{13}(\mathbf{k}) & -g_{23}(\mathbf{k}) & 0
\end{pmatrix},
\end{dmath}
where
\begin{align}
g_{12}(\mathbf{k})
&=
-\frac{1}{2}\,e^{\frac{i}{2}k_y}\,
\sin\!\left(\frac{\sqrt{3}}{2}k_x\right),\\
g_{13}(\mathbf{k})
&=
-\frac{1}{4i}\left[-e^{-ik_y}+e^{\frac{i}{2}(\sqrt{3}k_x+k_y)}\right],\\
g_{23}(\mathbf{k})
&=
\frac{1}{2}\,e^{-\frac{i}{4}(\sqrt{3}k_x+k_y)}\,
\sin\!\left(\frac{1}{4}(\sqrt{3}k_x-3k_y)\right).
\end{align}

\section{Dielectric function fit and convergence of the normal state results}\label{app:dielectric}

\begin{figure*}
    \centering
    \includegraphics[width=0.99\linewidth]{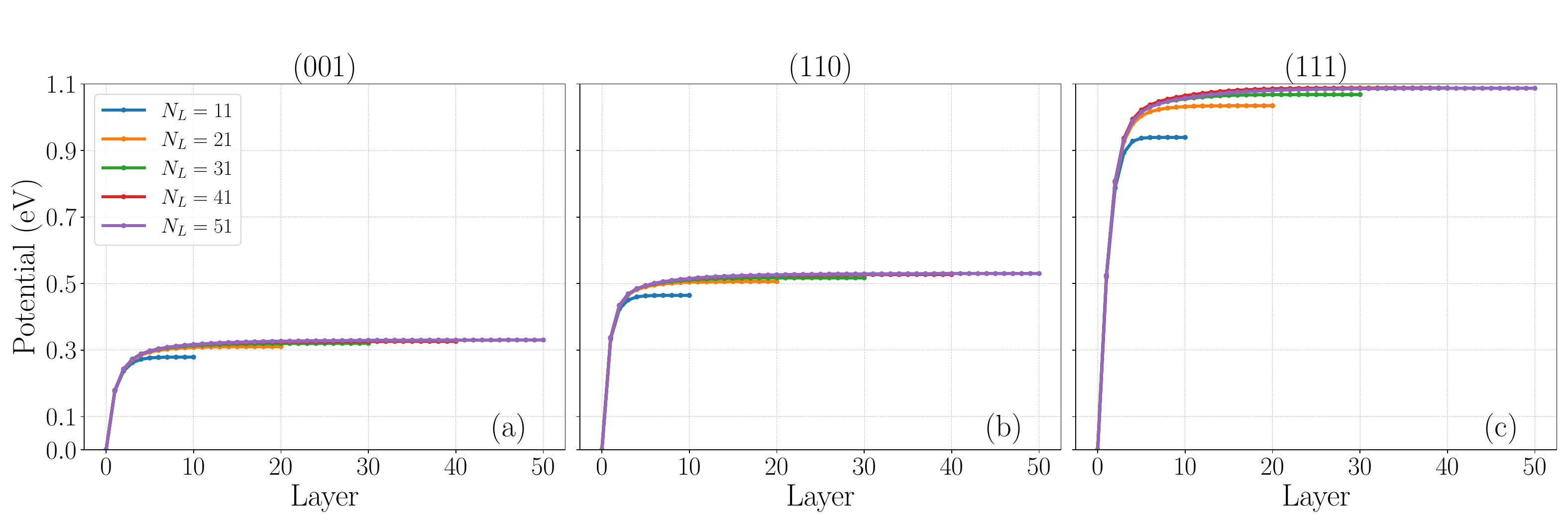}
    \caption{Electrostatic potential $\varphi$ as a function of the layer index for different slab thicknesses $N_L$ at $n_{2D}=1.6\times10^{14}$~cm$^{-2}$: (a) (001), (b) (110), and (c) (111) orientations.}
    \label{fig:potential_convergence}
\end{figure*}

Here we describe the fitting procedure adopted for the dielectric function $\varepsilon(F)$ entering Eqs.~\eqref{eq:maxwell}. 
We assume $\varepsilon(F)$ to be independent of both the crystallographic direction and the 2DEG density. 
Within this approximation, the screening properties at the interface are taken to be the same as in the bulk, and modeled through
\begin{equation}
    \varepsilon(F)=1+\frac{\chi_0}{\left(1+\frac{F^2}{F_{C}^2}\right)^{\gamma}},
\end{equation}
reproducing the low-temperature antiferroelectric response of KTO~\cite{dong2026strongly,bruneel2020electronic}. 
The parameters $F_C$, $\chi_0$, and $\gamma$ are determined through a fitting procedure. 
To reduce the number of free parameters, we restrict $\gamma$ to the values $\gamma=1/3$ and $2/5$, as commonly assumed in the literature~\cite{bruneel2020electronic,martinez20232deg,trama2023effect,copie2009towards}.

We define a benchmark electronic band structure for the (111) interface to be fitted against the ARPES data of Ref.~\cite{bruno20192d}. 
In Ref.~\cite{zhai2023large}, the band structure was reproduced using an effective potential ramp of the form
\begin{equation}
    \varphi_{\rm eff}(z)=\beta\left(1-e^{-z\frac{\alpha}{\beta}}\right),
    \label{eq:potential_fit_no_self}
\end{equation}
with $\alpha=0.4$~V, $\beta=0.97$~V, and $z$ the dimensionless layer index. 
This potential saturates for $z\to\infty$ and provides a good description of the experimental bands for $\mu=0.15$~eV above the band bottom.

Since the agreement between the confined band structure obtained with $\varphi_{\rm eff}$ and the experimental data is satisfactory in the first few layers, we proceed as follows. 
We extract the electronic density profile $n_z^{\text{Th}}$ over a 40-layer slab using Eq.~\eqref{eq:potential_fit_no_self} with $\mu=0.15$~eV, corresponding to a sheet density $n_{2D}^{\text{Th}}=1.6\times10^{14}$~cm$^{-2}$, as extracted from the ARPES-based band-structure analysis of Ref.~\cite{bruno20192d}. 
From this profile, we obtain the electric displacement
\[
    D_z=|e| \left( \sum_{i=1}^{40}n^{\text{Th}}_i-\sum_{i=1}^{z}n^{\text{Th}}_i \right),
\]
and, for a given $\varepsilon(F)$, the electric field $F_z$ from $D_z=\varepsilon_0 \varepsilon(F_z)\,F_z$. 
This is compared with the reference field derived from the effective potential,
\[
    F_z^{\text{Th}}=\frac{1}{\tilde{a}^\prime}\,\partial_z\varphi_{\rm eff}
    =\frac{\alpha}{\tilde{a}^\prime} e^{-z\frac{\alpha}{\beta}},
\]
where $\tilde{a}^\prime=3.988/\sqrt{3}$~\AA. 
The agreement is quantified through
\[
    \chi^2=\sum_{l=1}^{M}\frac{(F_l^{\text{Th}}-F_l)^2}{M(F_l^{\text{Th}})^2},
\]
which is minimized with respect to the parameters $F_C$, $\chi_0$, and $\gamma$, restricting the fit to the first $M=4$ layers, where $\varphi_{\rm eff}$ reliably reproduces the 2DEG bands.

The optimal parameters are $\gamma=1/3$, $\chi_0=3600$, and $F_C=443.7$~kV/m, consistent with the typical order of magnitude reported for related oxides~\cite{trama2023effect}.

Moreover, we studied the convergence of the electrostatic potential as a function of the total number of layers $N_L$. 
The goal is to verify that the slab is sufficiently thick to reproduce the bulk limit when the layer index approaches the boundary, $N_z \to N_L$.

Figure~\ref{fig:potential_convergence} shows the electrostatic potential profile obtained for different slab thicknesses at fixed sheet density $n_{2D}=1.6\times10^{14}$~cm$^{-2}$. 
Convergence to the bulk behavior is achieved for $N_L \gtrsim 41$. 
Accordingly, in all calculations presented in this work we set $N_L = 51$.

Finally, we quantify the residual electronic density far from the interface.
For this purpose, we define a tail region in the 51-layer slab as the set of layers $z\geq 20$, where the electrostatic potential is already close to its asymptotic bulk value. We then compute the residual sheet density
\begin{equation}
    n_{2D}^{\rm tail}=\sum_{z=20}^{51} n_z,
\end{equation}
and convert it into an average three-dimensional density as $
    \bar n_{3D}^{\rm tail}=n_{2D}^{\rm tail}/{L_{\rm tail}}$ ,
    $L_{\rm tail}=31 a^\prime$,
where $a^\prime$ is the out-of-plane lattice parameter for the corresponding
orientation. This quantity provides an orientation-independent estimate of the
residual bulk-like carrier density in the tail of the slab.
The results are reported in Table~\ref{tab:residual_tail_charge}. For all orientations and sheet densities considered, $\bar n_{3D}^{\rm tail}$ remains of the order of $10^{19}\,\mathrm{cm}^{-3}$ and shows only a weak dependence on the total imposed 2D density. This confirms that the additional charge mainly accumulates close to the interface, while the far end of the slab contains only a dilute residual tail. Together with the convergence of the electrostatic potential shown in Fig.~\ref{fig:potential_convergence}, this demonstrates that $N_L=51$ is sufficient for the normal-state quantities discussed in the main
text.

\begin{table}[t]
\centering
\caption{
Residual electronic density in the tail of the 51-layer slab. The tail region is defined as $z\geq 20$. Here $n_{2D}^{\rm tail}=\sum_{z=20}^{50} n_z$ is the residual sheet density contained in the tail, $L_{\rm tail}=31a^\prime$ is the corresponding physical thickness, and $\bar n_{3D}^{\rm tail}=n_{2D}^{\rm tail}/L_{\rm tail}$ is the average three-dimensional carrier density in this region.
}
\label{tab:residual_tail_charge}
\begin{tabular}{ccccc}
\hline
Orientation &
$n_{2D}$ &
$n_{2D}^{\rm tail}$ &
$L_{\rm tail}$ &
$\bar n_{3D}^{\rm tail}$ \\
&
$(10^{14}\,\mathrm{cm}^{-2})$ &
$(10^{13}\,\mathrm{cm}^{-2})$ &
$(\mathrm{nm})$ &
$(10^{19}\,\mathrm{cm}^{-3})$ \\
\hline\hline
(001) & 0.8 & 1.14 & 12.36 & 0.92 \\
(001) & 1.0 & 1.14 & 12.36 & 0.92 \\
(001) & 1.6 & 1.16 & 12.36 & 0.94 \\
\hline
(110) & 0.8 & 1.25 &  8.74 & 1.43 \\
(110) & 1.0 & 1.25 &  8.74 & 1.43 \\
(110) & 1.6 & 1.22 &  8.74 & 1.39 \\
\hline
(111) & 0.8 & 1.83 &  7.14 & 2.56 \\
(111) & 1.0 & 1.86 &  7.14 & 2.61 \\
(111) & 1.6 & 1.70 &  7.14 & 2.39 \\
\hline
\end{tabular}
\end{table}

\section{Mean field treatment of non-homogeneous superconductivity} \label{app:supercond} 

We provide details on the theoretical description of
inhomogeneous superconductivity.

Starting from a generic (screened) two-body interaction in real space
\begin{equation}
H_{\mathrm{int}}
=
\frac{1}{2}
\sum_{\sigma\sigma'}
\int d^3r\, d^3r'\;
\psi^\dagger_{\sigma}(\mathbf{r})
\psi^\dagger_{\sigma'}(\mathbf{r}')
U(\mathbf{r}-\mathbf{r}')
\psi_{\sigma'}(\mathbf{r}')
\psi_{\sigma}(\mathbf{r}),
\end{equation}
and expand the field operators in a localized (Wannier-like) orbital basis,
\begin{equation}
\psi_{\sigma}(\mathbf{r})
=
\sum_{\mathbf{R},\alpha}
\varphi_{\alpha}(\mathbf{r}-\mathbf{R})\,
c_{\mathbf{R}\alpha\sigma},
\end{equation}
where $\mathbf{R}$ labels lattice sites and $\alpha$ denotes local orbital
degrees of freedom.

Retaining only the dominant on-site, intra-orbital contribution yields the
effective local interaction
\begin{equation}
H_{\mathrm{int}}
\simeq
- U_{\mathrm{eff}}
\sum_{\mathbf{R},\alpha}
c^\dagger_{\mathbf{R}\alpha\uparrow}
c^\dagger_{\mathbf{R}\alpha\downarrow}
c^{\vphantom{\dagger}}_{\mathbf{R}\alpha\downarrow}
c^{\vphantom{\dagger}}_{\mathbf{R}\alpha\uparrow},
\end{equation}
with
\begin{equation}
U_{\mathrm{eff}}
=
\int d^3r\, d^3r'\;
|\varphi_{\alpha}(\mathbf{r})|^2\,
U(\mathbf{r}-\mathbf{r}')\,
|\varphi_{\alpha}(\mathbf{r}')|^2 .
\end{equation}
The effective coupling $U_{\mathrm{eff}}$ has dimensions of energy and depends
only on the microscopic interaction and on the localized orbital wavefunctions.
In the following we denote it simply by $U$. Since the system changes when
varying the crystallographic orientation, we adopt the same microscopic
convention for $U$ in all geometries.

Assuming local spin-singlet pairing, the interaction is decoupled at the
mean-field level as
    \begin{align}
- U\,
c^\dagger_{\mathbf{R}\alpha\uparrow}
c^\dagger_{\mathbf{R}\alpha\downarrow}
c^{\vphantom{\dagger}}_{\mathbf{R}\alpha\downarrow}
c^{\vphantom{\dagger}}_{\mathbf{R}\alpha\uparrow}
\;\rightarrow\;&
- U \left(
\Delta_{\mathbf{R}}^{*}
c^{\vphantom{\dagger}}_{\mathbf{R}\alpha\downarrow}
c^{\vphantom{\dagger}}_{\mathbf{R}\alpha\uparrow}
+
\Delta_{\mathbf{R}}
c^\dagger_{\mathbf{R}\alpha\uparrow}
c^\dagger_{\mathbf{R}\alpha\downarrow}
\right)
+ U |\Delta_{\mathbf{R}}|^2 ,
    \end{align}

where
\begin{equation}
\Delta_{\mathbf{R}}
=
\langle
c_{\mathbf{R}\alpha\downarrow}
c_{\mathbf{R}\alpha\uparrow}
\rangle ,
\end{equation}
and we assume the same order parameter for all orbitals.

\subsection*{Slab geometry and layer-resolved formulation}

We now consider slab geometries, where translational invariance is preserved
only in the plane. We decompose $\mathbf{R}=(\mathbf{r},z)$, with $\mathbf{r}$
the in-plane coordinate and $z$ the discrete layer index.

The superconducting order parameter becomes layer-resolved,
\begin{equation}
\Delta_z
=
\langle
c_{\mathbf{r}z\alpha\downarrow}
c_{\mathbf{r}z\alpha\uparrow}
\rangle ,
\end{equation}
while remaining uniform in-plane.

We introduce the partial Fourier transform
\begin{equation}
c_{\mathbf{r}z\alpha\sigma}
=
\frac{1}{\sqrt{\mathcal{A}}}
\sum_{\mathbf{k}}
e^{i\mathbf{k}\cdot \mathbf{r}}\,
c_{\mathbf{k}z\alpha\sigma},
\qquad
\sum_{\mathbf{k}}
=
\mathcal{A}\int\frac{d^2k}{(2\pi)^2},
\end{equation}
where $\mathcal{A}$ is the in-plane area.
With these conventions, the mean-field pairing term reads
\begin{equation}
H_{\mathrm{int}}
=
- U
\sum_{\mathbf{k}>0,z,\alpha}
\left(
\Delta_z^{*}
c_{\mathbf{k}z\alpha\uparrow}
c_{-\mathbf{k}z\alpha\downarrow}
+ \mathrm{h.c.}
\right)
+ 3U \mathcal{A} \sum_{z} |\Delta_z|^2 .
\end{equation}

Diagonalizing the full BdG Hamiltonian in particle-hole basis, we obtain
\begin{equation}
H =
\sum_{\mathbf{k}>0,n}
\lambda_{\mathbf{k}n}
\left(
\Gamma_{\mathbf{k}n}^\dagger \Gamma^{\vphantom{\dagger}}_{\mathbf{k}n}
+
\tilde{\Gamma}_{\mathbf{k}n}^\dagger
\tilde{\Gamma}_{\mathbf{k}n}^{\vphantom{\dagger}}
\right)
+ 3U \mathcal{A} \sum_{z} |\Delta_z|^2,
\end{equation}
where
$\lambda_{\mathbf{k}n} \geq 0$, up to constants.
Here $\Gamma_{\mathbf{k}n}$ and $\tilde{\Gamma}_{\mathbf{k}n}$ correspond to
the positive- and negative-energy BdG branches, respectively, and are related
by particle–hole symmetry.

The free energy is therefore
\begin{equation}
F =
- \frac{1}{\beta}
\sum_{\mathbf{k}>0,n}
\log\!\left[
2 + 2\cosh\!\left(\beta \lambda_{\mathbf{k}n}\right)
\right]
+ 3U \mathcal{A} \sum_{z} |\Delta_z|^2 ,
\label{eq:free_energy_app}
\end{equation}
with $\beta = 1/(k_B T)$.

It is convenient to rescale $F \rightarrow F/\mathcal{A}$ and convert the
momentum sum into an integral,
\begin{equation}
F =
3U \sum_{z} |\Delta_z|^2
-
\frac{\mathcal{P}}{\beta}
\sum_n
\int d^2\mathbf{k}\;
\log\!\left[
2 + 2\cosh\!\left(\beta \lambda_{\mathbf{k}n}\right)
\right],
\label{eq:free_energy_rescaled}
\end{equation}
where $\mathcal{P}$ includes the in-plane unit-cell area, the Jacobian
associated with the chosen coordinates, a factor $1/2$ from restricting to
$\mathbf{k}>0$, and the momentum-space density of states.

For the three crystallographic orientations we obtain
\begin{equation}
\mathcal{P}_{001} = \dfrac{1}{4\pi^2}, \quad
\mathcal{P}_{110} = \dfrac{\sqrt{2}}{4\pi^2}, \quad
\mathcal{P}_{111} = \dfrac{27}{8\pi^2}
\end{equation}

For the (111) orientation we use normal coordinates
$(\ell_x,\ell_y)$ defined by
\begin{equation}
k_x=\frac{\sqrt{3}}{2}\ell_x,
\qquad
k_y=\ell_y-\frac{\ell_x}{2}.
\end{equation}

Following the variational ansatz $\Delta_z=\Delta e^{-z/z_0}$, the free energy can be explicitly evaluated as
\begin{equation}
        F=G^2\left(\frac{\Xi(z_0)}{U}- \Gamma(z_0)\right),
        \label{eq:free_energy_compact}
\end{equation}
where $G=U\Delta$ is the superconducting gap opened at the first layer of the interface, and
\begin{align} 
    \Xi(z_0)&= 3 \frac{1-e^{-2N_L/z_0}}{1-e^{-2/z_0}},
    \label{eq:Xi}\\
    \Gamma(z_0) &= 
    \frac{\mathcal{P}}{2}
    \sum_n\int d^2\mathbf{k} 
    \tanh\left(\frac{\beta E_{n\mathbf{k}}}{2}\right)
    \frac{ f_{\mathbf{k}}^2(z_0)}{E_{n\mathbf{k}}}.
    \label{eq:Gamma}
\end{align}
By considering $
\pdv[2]{F}{G}=0$,
and
$\pdv{F}{z_0}=0$, we obtain the self-consisted values of the critical temperature.

For the phonon-mediated pairing mechanism, Eq.~\eqref{eq:Gamma} reduces to
\begin{equation}
    \Gamma(z_0)\approx
    \frac{\mathcal{P}}{2}\rho(z_0)
    \int_{-\omega_i}^{\omega_i}
    \frac{dE}{E}
    \tanh\left(\frac{\beta E}{2}\right),
    \label{eq:BCS_integral}
\end{equation}
where $\rho(z_0)$ is the weighted DOS, defined as
\begin{equation}
    \rho(z_0)=
    \sum_n\int d^2\mathbf{k} \,
    \delta(E_{n\mathbf{k}})\,
    f_{n\mathbf{k}}^2(z_0),
    \label{eq:dos_super}
\end{equation}
where $f_{n\mathbf{k}}^2(z_0)$ is the form factor weighing the projection of the normal-state eigenvectors across the slab~\cite{trama2025self}.
The integral in Eq.~\eqref{eq:BCS_integral} can be evaluated analytically, yielding
\begin{equation}
    \Gamma(z_0)=
    \frac{\mathcal{P}}{2}
    \rho(z_0)
    \Phi\!\left(\frac{\beta\omega_i}{2}\right),
\end{equation}
where $\Phi(x)\approx 2\log(2.27x)$.

The condition $T=T_c$  reads
\begin{equation}
    U=
    \frac{2}{\mathcal{P}\Phi(\beta\omega_i/2)}
    \min_{z_0}
    \left(
    \frac{\Xi(z_0)}{\rho(z_0)}
    \right),
\end{equation}
or, equivalently,
\begin{equation}
\begin{cases}
    k_B T_c = 1.13\,\omega_i
    \exp\!\left(-\dfrac{\tilde{U}}{2U}\right), \\[6pt]
    \tilde{U} =
    \dfrac{2}{\mathcal{P}}
    \displaystyle\min_{z_0}
    \left(
    \dfrac{\Xi(z_0)}{\rho(z_0)}
    \right).
\end{cases}
\label{eq:phonon_minimisation}
\end{equation}

\subsection*{Istantaneous paring mechanism}

In the instantaneous pairing scenario, the full electronic bandwidth contributes to the pairing, albeit with a logarithmic predominance of states close to the Fermi surface. This scenario is particularly relevant when the chemical potential approaches regions of strong band reconstruction or band folding. 
Minimizing with respect to $z_0$ gives 
\begin{equation}
    \frac{\Xi'(z_0)}{U}-\Gamma'(z_0)=0,
    \label{eq:free_energy2}
\end{equation}
where the prime denotes differentiation with respect to $z_0$. Combining this with the instability condition $\partial^2 F/\partial G^2=0$, we obtain the coupled equations determining $U$ and $z_0$ at $T=T_c$:
\begin{equation}
\begin{cases}
    \partial_{z_0}\log\left(\dfrac{\Xi}{\Gamma}\right)=0, \\[6pt]
    U=\dfrac{\Xi}{\Gamma}.
\end{cases}
\label{eq:istantaneous_minimisation}
\end{equation}

\begin{figure*}[t]
    \centering
    \includegraphics[width=0.97\textwidth]{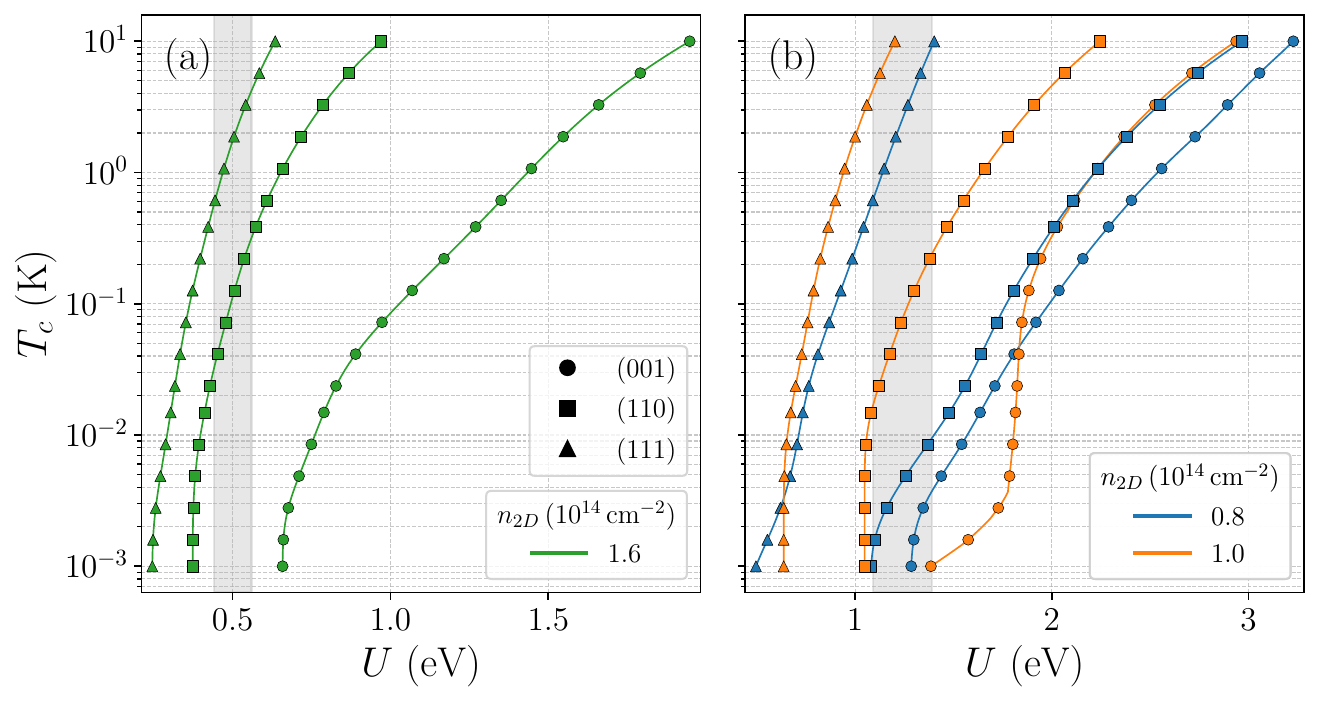}
    \caption{
    Critical temperature $T_c$ as a function of the interaction strength $U$ in the instantaneous paring scenario for (a) $n_{2D}=1.6\times10^{14}$~cm$^{-2}$, and (b) $n_{2D}=0.8\times10^{14}$~cm$^{-2}$ (blue) and $n_{2D}=1.0\times10^{14}$~cm$^{-2}$ (orange). 
    Different markers denote different crystallographic orientations: circles (001), squares (110), and triangles (111).
    The gray area highlights the region of $U$ which are able to reproduce the experimental order of magnitude of $T_c$.}
    \label{fig:comparison.pdf}
\end{figure*}

In Fig.~\ref{fig:comparison.pdf} we show the resulting critical temperature $T_c$ as a function of the interaction strength $U$ within this cutoff-free pairing model. Also in this case, allowing for an inhomogeneous order parameter reproduces the hierarchy found in the phonon-mediated scenario.

In contrast to the phonon-mediated case, the dependence of $T_c$ on $U$ is not purely exponential at small interaction strengths. In particular, the (001) interface displays a threshold-like behavior, which disfavors the onset of superconductivity. For sufficiently large $U$, however, the curves recover an exponential dependence, consistent with the asymptotic behavior obtained using the phonon-mediated pairing and discussed in the Appendix \ref{app:details_z0}. The amplitudes, on the other hand, differ among the three orientations, as they are determined solely by the normal-state electronic structure.

We also observe that there exists a range of interaction strengths yielding critical temperatures of the correct experimental order of magnitude for all three orientations. In Fig.~\ref{fig:comparison.pdf}(a), the shaded region highlights this interval at fixed density. In panel (b), the experimentally observed hierarchy is recovered only when different carrier densities are considered for the three interfaces.

\begin{figure*}
    \centering
    \includegraphics[width=0.99\linewidth]{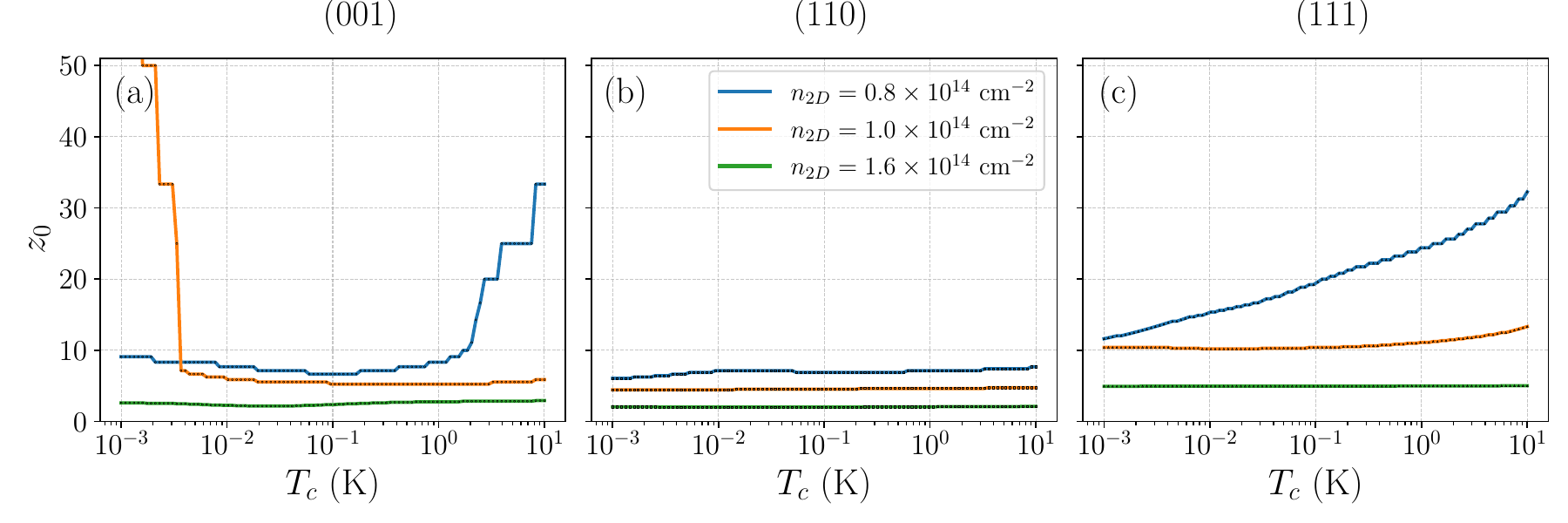}
    \caption{Superconducting penetration length $z_0$ as a function of the imposed critical temperature $T_C$ for (a) (001), (b) (110), and (c) (111) directions, and for different benchmark values of $n_{2D}$ in the istantaneous paring scenario.}
    \label{fig:tc_z0_full_range}
\end{figure*}

We now comment on the spatial extension of the superconducting order parameter across the slab.
In the phonon-mediated case (see Fig.~\ref{fig:BCS_all.pdf}), there is a direct correlation between the spatial extension of the 2DEG and $z_0$, since most of the LDOS at the Fermi level is localized within the 2DEG region.
On the other hand, for the instantaneous pairing scenario, the superconducting penetration length $z_0$ at which the instability occurs depends explicitly on the chosen critical temperature. In Fig.~\ref{fig:tc_z0_full_range} we show the evolution of $z_0$ as a function of the imposed $T_C$. The behavior is non-trivial. In particular, for the (001) direction there are regimes in which the most favorable solution corresponds to superconductivity extending throughout the entire slab. This behavior also depends on the carrier density.
Such non-trivial trends originate from the structure of the available 2DEG states. Reaching an instability at a given $T_C$ requires either a sufficient density of states near the interface or a stronger attractive pairing interaction. Alternatively, the instability can be supported by involving bulk states farther from the interface.

For (111) direction at low density, the increase of $z_0$ with $T_C$ shows the necessity of involving bulk states, whereas at higher density the electronic states localized near the interface are sufficient to sustain superconductivity without a significant extension into the bulk. 
The (110) direction, on the other hand, appears to be the most confined case, as it consistently involves contributions from folded bands at the Brillouin-zone boundary, predominantly localized in the first layer, where the LDOS exhibits a pronounced peak.

The analysis above clarifies how the instability optimizes its spatial profile through the self-consistent determination of $z_0$. 

\section{Impact of confinement of superconductivity and comparison between the two scenarios}\label{app:details_z0}

For both the retarded and the instantaneous pairing mechanisms discussed in the main text, we minimized Eqs.~\eqref{eq:phonon_minimisation} and~\eqref{eq:istantaneous_minimisation}. Here we present benchmark results for $\tilde{U}(z_0)$ and $U(z_0)$ in the two cases, together with the self-consistent values of $z_0$ obtained for the instantaneous interaction. This analysis allows us to further comment on the assumption of homogeneous pairing and on the necessity of going beyond this limitation.

\begin{figure*}
    \centering
    \includegraphics[width=0.93\textwidth]{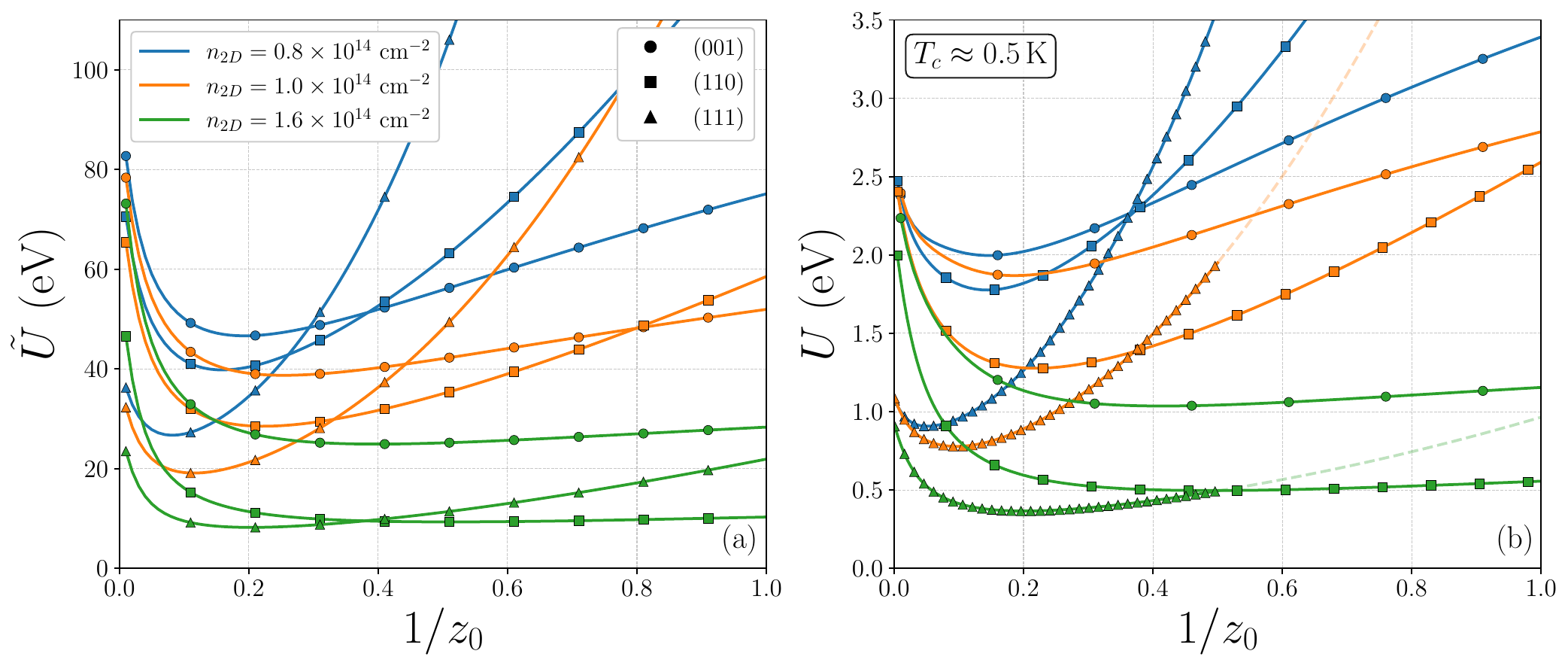}
    \caption{(a) Exponential factor $\tilde{U}$ as a function of $1/z_0$. The limit $1/z_0=0$ corresponds to purely homogeneous superconductivity. (b) Superconducting coupling strength $U$ as a function of $1/z_0$ for a benchmark $T_c \approx 0.5$~K. For the (111) direction, dashed lines are fitted curves used only to interpolate missing points and guide the eye. In both panels, circles denote (001), squares (110), and triangles (111) directions, while colors correspond to different $n_{2D}$.}
    \label{fig:Utilde_vs_z0}
\end{figure*}

\begin{figure*}
    \centering
    \includegraphics[width=0.96\linewidth]{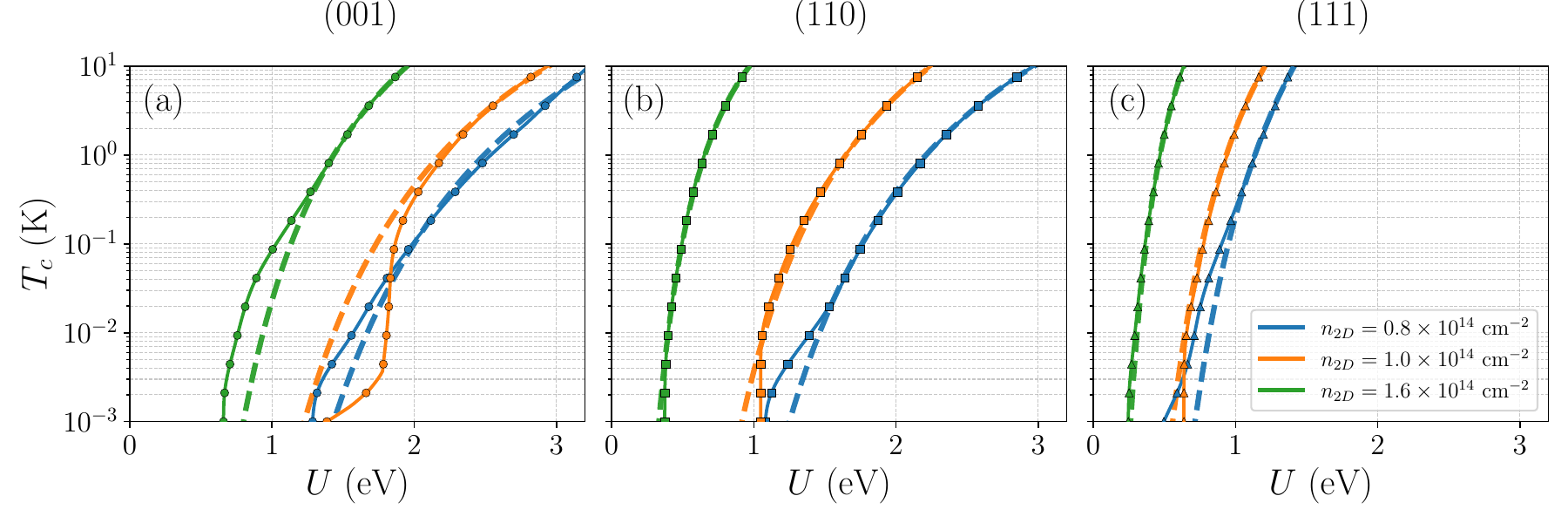}
    \caption{Critical temperature $T_C$ for (a) (001), (b) (110), and (c) (111) directions, evaluated within the instantaneous interaction mechanism (solid lines and markers). The dashed lines correspond to an exponential fit $T_C \sim \omega_i , \mathrm{exp}(-\tilde{U}/2U)$, where $\tilde{U}$ is obtained self-consistently from the phonon-mediated pairing framework and $\omega_i$ is used as a fitting parameter in the large-$U$ regime.}
    \label{fig:fit_phonons}
\end{figure*}

In Fig.~\ref{fig:Utilde_vs_z0} we show the exponential factor $\tilde{U}$ as a function of $1/z_0$ for the phonon-mediated pairing mechanism, and the superconducting coupling strength $U$ as a function of $1/z_0$ for a fixed benchmark critical temperature $T_c \approx 0.5$~K in the instantaneous case. 
In both cases we observe pronounced minima at finite values of $1/z_0$, corresponding to a preferred inhomogeneous solution. 
Interestingly, if one enforces a homogeneous solution in the instantaneous pairing mechanism, the values of $U$ required to obtain $T_c \approx 0.5$~K are almost independent of $n_{2D}$. This indicates that the dominant contribution to superconductivity would not originate from the 2DEG, but rather from bulk states, an assumption that is physically implausible in the present context.

The difference between the predicted critical temperatures between the two scenarios stems from the distinct functional dependence of the instability condition on the pairing strength in the two mechanisms. While in the phonon-mediated case the transition temperature follows a purely exponential BCS-like form, in the instantaneous interaction the dependence of $T_C$ on $U$ deviates from a simple exponential at small coupling. This deviation is directly linked to the self-consistent determination of the optimal penetration length $z_0$, which modifies the effective phase space available for pairing.
In Fig.~\ref{fig:fit_phonons} we make this comparison explicit. We plot the critical temperature obtained within the instantaneous mechanism as a function of $U$, and compare it with an exponential form $T_C \sim \omega_i \exp(-\tilde{U}/2U)$, where $\tilde{U}$ is taken from the phonon-mediated framework and $\omega_i$ is used as a fitting parameter in the large-$U$ regime.

For sufficiently large interaction strengths, all orientations recover an exponential dependence, indicating that the two mechanisms share the same asymptotic behavior once the optimal spatial profile is selected. The residual differences between the (001), (110), and (111) interfaces are then encoded in the prefactors, set by the normal-state electronic structure and the distribution of low-energy states across the slab.
At small $U$, deviations from exponential behavior emerge, particularly for the (001) orientation, where a threshold-like onset of superconductivity appears. This highlights the crucial role of the interplay between confinement, density of states, and the spatial extent of the order parameter in the weak-coupling regime.

\twocolumngrid

\bibliography{Bib}
\bibstyle{unsrt}

\end{document}